\documentclass[pra,twocolumn,amsmath,amssymb,superscriptaddress,english,aps,prd,longbibliography]{revtex4-2}
\usepackage[english]{babel}
\usepackage[utf8]{inputenc}
\usepackage{dsfont} 
\usepackage{siunitx}
\usepackage{subcaption}
\usepackage{algorithm}
\usepackage{algpseudocode}
\usepackage{multirow}
\usepackage[unicode=true,pdfusetitle, bookmarks=true,bookmarksnumbered=false,bookmarksopen=false,
 breaklinks=false,pdfborder={0 0 1},backref=false,colorlinks=true]
 {hyperref}
\usepackage{physics}
\usepackage[capitalize]{cleveref}
\usepackage{graphicx}

\newcommand\remove[1]{}

\makeatletter
\renewcommand{\selectlanguage}[1]{} 
\newcommand\moso{M\o{}lmer-S\o{}rensen gate}

\newcommand{\bigO}[1]{\mathcal{O}(#1)}
\makeatother

\begin{document}

\author{Susanna Kirchhoff}
\affiliation{Peter Grünberg Institute for Quantum Computing Analytics (PGI-12),
Forschungszentrum Jülich, 52425 Jülich, Germany}
\affiliation{Theoretical Physics, Universität des Saarlandes, 66123 Saarbrücken, Germany}
\author{Frank K. Wilhelm}
\affiliation{Peter Grünberg Institute for Quantum Computing Analytics (PGI-12),
Forschungszentrum Jülich, 52425 Jülich, Germany}
\affiliation{Theoretical Physics, Universität des Saarlandes, 66123 Saarbrücken, Germany}
\author{Felix Motzoi}
\affiliation{Peter Grünberg Institute for Quantum Control (PGI-8),
Forschungszentrum Jülich, 52425 Jülich, Germany}
\affiliation{Institute for Theoretical Physics, University of Cologne, 50937 Cologne, Germany}

\title{Correction formulas for the \moso{} under strong driving}

\begin{abstract}
    The \moso{} is a widely used entangling gate for ion platforms with inherent robustness to trap heating. The gate performance is limited by coherent errors, arising from the Lamb-Dicke (LD) approximation and sideband errors. Here, we provide explicit analytical formulas for errors up to fourth order in the LD parameter, by using the Magnus expansion to match numerical precision, and overcome significant, orders-of-magnitude underestimation of errors by previous theory methods. We show that fourth order Magnus expansion terms are unavoidable, being in fact leading order in LD, and are therefore critical to include for typical experimental fidelity ranges. We show how these errors can be dramatically reduced compared to previous theory by using analytical renormalization of the drive strength, by calibration of the Lamb-Dicke parameter, and by the use of smooth pulse shaping. 
\end{abstract}

\maketitle

\section{Introduction}\label{sec:Introduction}
A ubiquitous aspect of quantum processor design is the need to provide a strong contrast, or on/off ratio, between entanglement coupling vs.~idling and local single-qubit gates. To this effect, the standard approach is to introduce new coupling elements, whose auxiliary Hilbert space provides a means through which entanglement can temporarily be mediated, for example via intermediate cavity bus \cite{paik_experimental_2016,heya_cross-cross_2021,li_experimental_2024}, non-linear resonators \cite{barends_diabatic_2019, reagor_demonstration_2018, negirneac_high-fidelity_2021}, motional trapping modes \cite{schmidt-kaler_realization_2003, benhelm_fault-tolerant_2008, schafer_fast_2018, mehta_fast_2019, clark_high-fidelity_2021, katz_body_2022, weber_robust_2024}, intermediary spins \cite{fei_mediated_2012, casanova_arbitrary_2017}, etc.

To fully leverage these coupling elements, and avoid the onset of decoherence, it is desirable to turn on the effective coupling between qubits as strongly as possible, for example through the use of strong driving fields. The prototypical example of this is the so-called \moso{} (MSG). This technique was first introduced in \cite{sorensen_quantum_1999} to be resilient against trap heating within the Lamb-Dicke and weak driving perturbative approximations \cite{lishman_trapped-ion_2020,arrazola_pulsed_2018,bermudez_robust_2012,cohen_multi-qubit_2015,jia_angle-robust_2023,valahu_quantum_2022,kang_batch_2021}. And yet, it is important to recognize that wide adoption of the gate largely occurred after a further advancement, whereby strong driving could be reinstituted at the cost of the collective ion motion being temporarily activated before returning to its initial state \cite{sorensen_entanglement_2000}. 

In the wider context of trapped ion quantum computing \cite{schindler_quantum_2013,haffner_quantum_2008,brown_co-designing_2016}, the platform has blossomed to one of the leading candidates for housing quantum information, boasting long coherence times around \SIrange{1}{50}{\second} \cite{kaufmann_scalable_2017,harty_high-fidelity_2016}, and infidelities for two-qubit gates on the order of \num{1e-3} at a gate duration of \SIrange{1.6}{300}{\micro\second} \cite{schafer_fast_2018,mehta_fast_2019,clark_high-fidelity_2021,benhelm_fault-tolerant_2008}. 

Further advancements hinge on proper theoretical understanding of the effect of strong driving on the operation with proper accounting of the excitation of the oscillator states, and potentially reduced gate durations. As with the oscillator-based couplings in other platforms, there are limited toolboxes available for tracking the dynamical effects of strong driving, the two most common being the Dyson and Magnus expansions. In the \moso{}, the standard argument, dating back 25 years to Ref.~\cite{sorensen_entanglement_2000}, argues that, within the Lamb-Dicke approximation, an effective Hamiltonian in the Magnus expansion leads to only a single order being nonzero and accounting for the entanglement. The gate itself is second order in the Lamb-Dicke parameter $\eta$, while the error is fourth order in $\eta$, appearing in the same Magnus order, and so dictates the ultimate error scaling of the gate.

In this paper, we demonstrate that, on the contrary, there are multiple non-zero $\mathcal O(\eta^2)$ terms in the effective Hamiltonian that contribute even more significantly to the gate error. These errors are higher order in the Rabi frequency, but since this cannot be considered a small parameter (compared especially to the sideband detuning), these are fundamental to the functioning of the gate. We show that, in general, neglecting these terms leads to large detectable errors and deviations from experiment. These translate for typical parameters to circa \SI{10}{\percent} Rabi frequency mismatch, and, without in-situ recalibration, 10-100 times larger error than theoretically predicted. 

We provide analytical expressions for the leading orders of those terms and give a correction term for the drive strength $\Omega$ that can partially compensate for the error caused by those terms. Our model results in a bound on the gate infidelity of about $10^{-3}$ for typical parameters, which matches well with experimental fidelity measurement results \cite{ballance_high-fidelity_2016,schafer_fast_2018,mehta_fast_2019, gerster_experimental_2022,moses_race-track_2023, saner_breaking_2023}.
We also motivate analytically and show numerically how it is – depending on the parameter regime – of advantage to use pulse shaping, to improve the gate fidelity closer to $10^{-4}$. This can help to approach fidelities needed for fault-tolerant quantum computing \cite{bermudez_assessing_2017, zarantonello_robust_2019}.

The remaining structure of the paper is as follows:
In Section \ref{sec:Model}, we (A) review  the Magnus and the Dyson methods to calculate the propagator of the composite system; (B) discuss the effect of strong driving on the convergence the Magnus expansion; and (C) define the system Hamiltonian and its expansion into Taylor- and Fourier coefficients.
Next, in  Section \ref{sec:Resonance Conditions}, we explain a way to calculate arbitrary orders of the Magnus expansion for the given system, and investigate resonance conditions for turning on and off certain transitions. 
In Section \ref{sec:Error terms}, we provide analytical expressions for the different terms of the Magnus expansion, for the typical case of a rectangular drive amplitude, giving correction formulas accounting for higher Magnus order errors. We also evaluate numerically the importance of different system parameters on those errors. In Section \ref{sec:Pulse Shaping}, we further provide analytical expressions and evaluate terms numerically, this time for shaped drive amplitudes.
We finally summarize in Section \ref{sec:Summary}.

\crefalias{section}{appendix}

\section{Model and Methods}\label{sec:Model}

\begin{figure*}[ht]
    \centering
    \includegraphics[width=\linewidth]{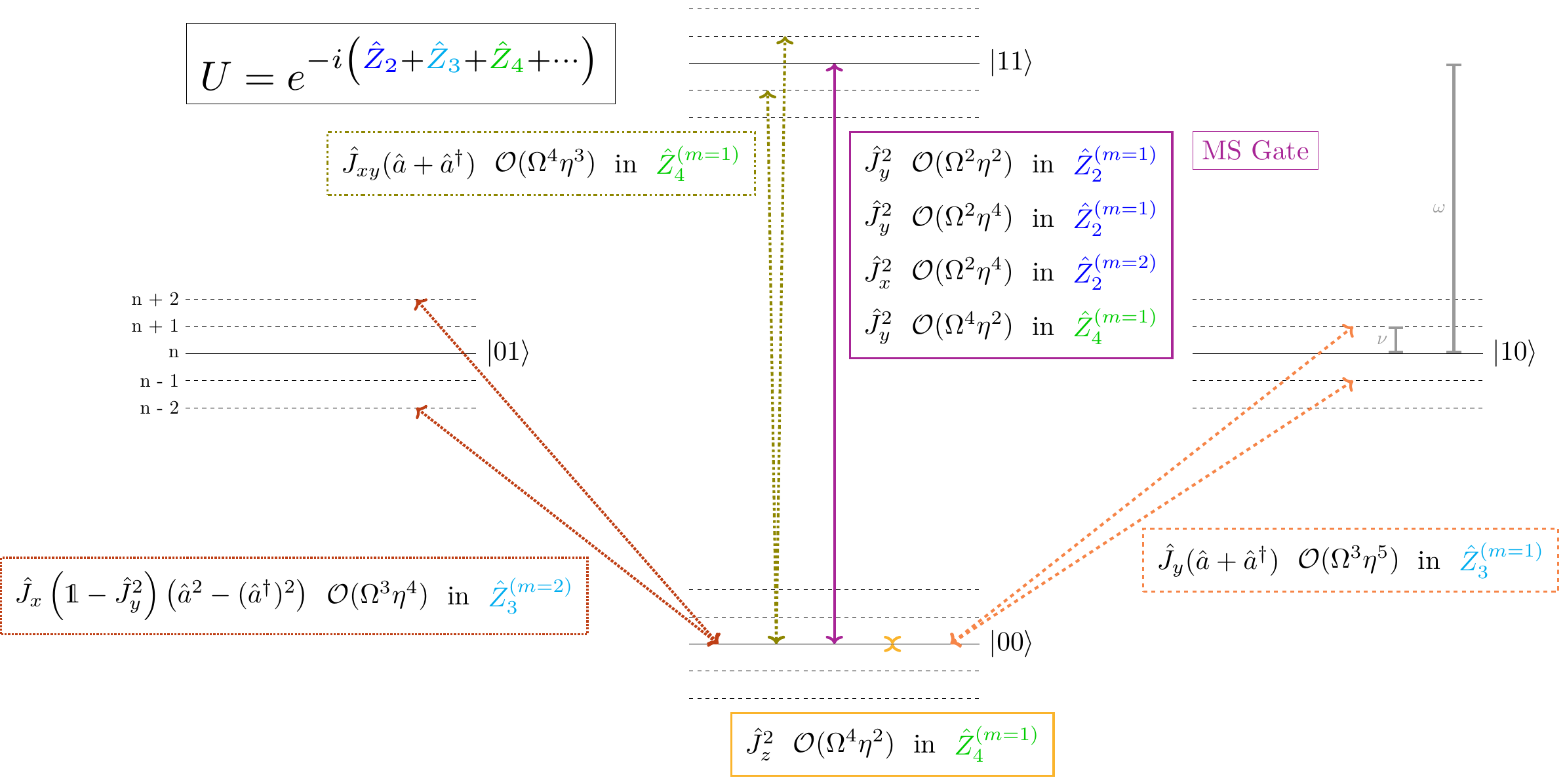}
    \caption{Energy level diagram for the system in the interaction picture, with transitions connected to the $\ket{00}$ state being shown. The straight black lines depict the energy of the collective qubit state, and the dashed black lines depict coupled qubit and motional mode states. The solid pink arrow between $\ket{00}$ and $\ket{11}$ shows the transitions which create the entangling gate. Those errors can be corrected by adjusting the drive amplitude. The dashed-dotted green arrows left to the solid arrows show transitions that create an entangling operation in the qubit subspace, but additionally cause leakage into different motional mode states. The term proportional to $\hat J_z^2$ (yellow, at $\ket{00}$) only adds a phase. The dashed orange lines on the right-hand side denote first-sideband ($m=1$) off-resonant error terms, and the dotted red lines on the left-hand side denote the second-sideband ($m=2$) off-resonant error terms. Those terms can be minimized by time-shaping the drive amplitude. The full terms are listed in \cref{tab:Error budget}. The lines denoting the transitions are just a sketch to give an intuition on the impact of each term, so not all potential transitions are shown.
		}
    \label{fig:Energy level diagram}
\end{figure*}

\subsection{Time propagation}
We first discuss the general methodology for obtaining the time propagation of our driven system. We want to first emphasize that, for a system such as the \moso{}, the drive dependence cannot in general be described with perturbation theory, since it may be larger than other relevant energy scales. Therefore, where multiple small parameters may exist, these different contributions should be properly accounted to avoid early truncation.

To this end, we describe the two most common expansions to obtain the propagator $\hat U(T)$ of a quantum system that gives the time evolution between time $0$ and time $T$.

The Dyson expansion \cite{sakurai_modern_1994} gives the propagator as 
\begin{equation}
    \label{eq:Dyson}
    \hat U(T) = \mathds{1} + \sum_{k=1}^\infty \hat P_k(T),
\end{equation}
where each term of the Dyson sum can be written as a time-ordered integral
\begin{equation}
    \label{eq:Dyson terms}
   \hat P_k(T) = \left(-\frac{i}{\hbar}\right)^k\int_0^T dt_1\cdots \int_0^{t_{k-1}}dt_k \hat H(t_1)\cdots \hat H(t_k).
\end{equation}
This expansion is mostly used for systems with a small perturbative interaction Hamiltonian.

An expansion with somewhat better convergence properties is the Magnus expansion \cite{magnus_exponential_1954}. It is exact for systems that are bosonic, and often converges quickly for fast oscillating systems, as is the case here. The Magnus propagator \cite{warren_effects_1984}
 \begin{equation}
     \label{eq:Magnus}
     \hat U(T) = e^{-i\sum_{k=1}^\infty\hat Z_k(T)}
 \end{equation}
 is calculated with a time-independent dimensionless effective Hamiltonian $\sum_{k=1}^\infty\hat Z_k(T)$, which gives after time $T$ the same dynamics as the original Hamiltonian. The effective Hamiltonian is the sum of the different orders of the Magnus expansion. The first three orders are
\begin{equation}
     \label{eq:Z def}
\begin{aligned}
		\hat Z_1 &= \frac{1}{\hbar}\int_0^T dt \hat H(t) ,\\
		\hat Z_2 &= \frac{i}{2\hbar^2} \int_0^T dt \int_0^t d\tau \comm{\hat H(t)}{\hat H(\tau)},\\
		\hat Z_3 &= -\frac{1}{6\hbar^3}\int_0^T dt \int_0^t d\tau \int_0^{\tau} d\tau' \left(\comm{\comm{\hat H(t)}{\hat H(\tau)}}{\hat H(\tau')}\right.\\&+\left.\comm{\hat H(t)}{\comm{\hat H(\tau)}{\hat H(\tau')}}\right).
 \end{aligned}
\end{equation}
There are also formulas for higher orders \cite{blanes_magnus_2009}, which we do not explicitly write down here due to their complexity.

Salzmann \cite{salzman_alternative_1985} gives instead a formula which expresses the orders of the Magnus expansion in terms of different orders of the Dyson expansion. Using his result, one can easily show that if $i\hat P_1 = \hat Z_1 = 0$, i.e., if the perturbation is unbiased, the relations between Magnus and Dyson expansion are
\begin{equation}
    \label{eq:Magnus terms for P1=0}
\begin{aligned}
    \hat Z_2 &= i\hat P_2,\\
    \hat Z_3 &= i\hat P_3,\\
    \hat Z_4 &= i\left(\hat P_4 - \frac{1}{2} \hat P_2^2\right),\\
    \hat Z_5 &= i\left(\hat P_5 - \frac{1}{2}\left(\hat P_2\hat P_3+\hat P_3\hat P_2\right)\right).
\end{aligned}
\end{equation}
Clearly, \cref{eq:Magnus terms for P1=0} shows that the first two nonzero orders of the Magnus expansion equal the first nonzero orders of the Dyson expansion, except for a phase. Starting at fourth order, the terms for Dyson and Magnus expansion differ. This connection between Magnus and Dyson expansion is quite powerful, because, unlike the terms of the Magnus expansion, the terms of the Dyson expansion do not involve nested commutators.
In the following sections, we calculate the Magnus expansion using \cref{eq:Magnus terms for P1=0}, allowing us to efficiently and analytically track the role of different errors.
Note that while the Dyson expansion gives an approximated propagator, the Magnus expansion gives an effective time-independent Hamiltonian, which is then used to calculate the propagator.

\subsection{Perturbations to driven harmonic oscillator}

To underscore the existence of error mechanisms arising from low-perturbation yet higher-order scattering processes under strong driving, we first show how they emerge under a generic driven Hamiltonian. 

We consider a perturbative coupling between a coupler element (here harmonic oscillator) and another (computational) subsystem, and work in an appropriate rotating frame or interaction picture. This encompasses the Hamiltonian of the \moso{}, but also coupling mechanisms for a wide variety of other fast two-qubit gates \cite{paik_experimental_2016,heya_cross-cross_2021,li_experimental_2024,barends_diabatic_2019, reagor_demonstration_2018, negirneac_high-fidelity_2021,schmidt-kaler_realization_2003, benhelm_fault-tolerant_2008, schafer_fast_2018, mehta_fast_2019, clark_high-fidelity_2021, katz_body_2022, weber_robust_2024} and readout \cite{boissonneault_improved_2010, reed_high-fidelity_2010, jerger_dispersive_2024} mechanisms.

We start with generically expanding the interaction in orders of a small parameter $\lambda$, 
\begin{equation}\label{eq:Hamiltonian general approx}
\hat H =   \hat H_0(t) + \lambda\hat H_1(t) + \mathcal{O}(\lambda^2),
\end{equation}
which acts on the composite space $\mathcal{H}_\text{HO}\otimes\mathcal{H}_\text{qubits}$. 
Now, we consider the effect of different resonance offsets ($\omega_k$) accompanying the different perturbation orders of the Hamiltonian ($\hat H_k$), or more precisely
\begin{equation}\label{eq:DHO}
		\hat{H}_k(t)=\hbar \Omega(t) \hat a^k e^{-i\omega_k t}\otimes \hat O_{k}+ H.c.,
\end{equation}
where $\Omega$ is the drive amplitude, which we cannot always assume is smaller than the resonance offsets $\omega_k$ due to strong driving, $\hat O_k$ are operators acting on the qubit subspace, and $\hat a, \hat a^\dagger$ are the annihilation and creation operators, respectively. 
Note that we have fixed the particular case of a common drive amplitude and specific powers of the harmonic oscillator operators, but this is simply done for concreteness and readability.

In the first order Magnus expansion, we have $\hat Z_1=\frac{1}{\hbar}\int_0^T dt \hat H(t)$ corresponding to oscillating (off-resonant) terms. Often these are enforced to integrate to zero, but not strictly needed for our purposes in this subsection.

The second order of the Magnus expansion is given by
\begin{equation}
\hat Z_2 = \frac{i}{2\hbar^2}\int_0^T dt \int_0^t d\tau \comm{\hat H(t)}{\hat H(\tau)},
\end{equation}
where the commutator can be broken down as
\begin{equation}
	\label{eq:general comm Z2}
	\begin{aligned}
		\comm{\hat H(t)}{\hat H(\tau)} &=  \comm{\hat H_0(t)}{\hat H_0(\tau)} +\lambda^2\comm{\hat H_1(t)}{\hat H_1(\tau)} \\
         &+\lambda\comm{\hat H_1(t)}{\hat H_0(\tau)}+\lambda\comm{\hat H_0(t)}{\hat H_1(\tau)} + \dots
	\end{aligned}
\end{equation}
The important point to note is that the first line of \cref{eq:general comm Z2} corresponds to resonant driving while the second does not, since the zeroth and first order frequency offsets are non-commensurate. In particular, in later sections the $\lambda^2$ term will correspond to the MSG interaction.

The third order of the Magnus expansion similarly contains a sum of commutators as in \cref{eq:Z def},
\begin{equation}
		\comm{\comm{\hat H(t)}{\hat H(\tau)}}{\hat H(\tau')}+\comm{\hat H(t)}{\comm{\hat H(\tau)}{\hat H(\tau')}},
\end{equation}
 but which are always non-resonant, since there is an odd number of factors. Unlike in the first order, these are generally not calibrated to integrate to zero, but these errors will still be relatively small.

Meanwhile, going to the fourth order $\hat Z_4$, the Magnus expansion will introduce commutators of the form
\begin{equation}
		\lambda^2\comm{\comm{\comm{\hat H_0(t)}{\hat H_1(\tau)}}{\hat H_0(\tau')}}{\hat H_1(\tau'')},
\end{equation}
and other permutations thereof, that is where we have an even number of $H_0$ and $H_1$ factors.

We see that we have a resonant coupling interaction of the same perturbation order as the one arising from $\hat Z_2$. Because we drive relatively strongly, the fact that the interaction is fourth order in $\Omega$ does not in general suppress the transition.

The general takeaways are twofold. The first is that the perturbation order of the Hamiltonian need not be linked with increasing expansion orders when the driving itself is non-perturbative. This is especially true in many-level systems where spurious resonances may arise. The second takeaway is that the expansion is non-monotonic, e.g.,~with the error increasing for the first four orders before eventually reducing again.

Therefore, in this general case, it is insufficient to terminate the series at low order. \cref{fig:Energy level diagram} shows the parameter scaling of errors in the Magnus expansion for the MSG Hamiltonian studied in this paper, where $\lambda=\eta$ is the Lamb-Dicke factor. Such higher-order processes must be carefully dealt with whenever coupler elements are driven strongly to accelerate entangling operations between subsystems.

\subsection{MSG System Hamiltonian}\label{sec:System}
We model the trapped ion system as two qubits coupled to a harmonic oscillator \cite{wineland_experimental_1998}.
Despite the capability of the \moso{} to operate pairwise on ion chains of arbitrary length, we focus for the sake of simplicity in this paper on two-qubit gates, without loss of generality.
The Hamiltonian of the uncoupled two-qubit system is $H_q = \hbar \omega_q\hat J_z$ with qubit transition frequency $\omega_q$ and a collective spin operator defined by
\begin{equation}\label{eq:def J}
		\hat J_i = \frac{1}{2}\left(\mathds{1}\otimes\hat \sigma_i + \hat \sigma_i \otimes \mathds{1}\right),
\end{equation}
where $\sigma_i$ (for $i=x,y,z$) are the Pauli operators.

Along the z-axis of the trap, the ions are confined in a harmonic effective potential. Their motion can be described in terms of motional modes \cite{james_quantum_2000}. 
When there are two ions, the two available modes are the so-called breathing mode and center-of-mass mode.
In the following, we take only one of the modes into account, which is a valid assumption as long as the drive detuning is close to the frequency of the mode \cite{sorensen_entanglement_2000} (see \cref{sec:Magnus term condition}).
In the Hamiltonian, the motional mode is modeled as a harmonic oscillator with frequency $\nu$, and lowering and raising operator $\hat a,\,\hat a^\dagger$, respectively.

The interaction between the systems is invoked by a classical light field with drive envelope $\Omega(t)$ and wave number $k$ along the trap x-axis \cite{leibfried_quantum_2003}. 
The entangling operation of the \moso{} is generated by driving two laser frequencies at the same time \cite{sorensen_entanglement_2000}. When each laser is detuned by $\pm \delta$ from the qubit frequency $\omega_q$, the Hamiltonian can be written as
\begin{equation}
    \label{eq:Hamiltonian lab frame}
    \hat H = \hbar \omega_q\hat J_z + \hbar \nu \hat a^\dagger \hat a + 2 \hbar\Omega(t)\sum_{\mu=\pm1}\cos(k\hat q - (\omega_q +\mu\delta )t) \hat J_x,
\end{equation}
with $\hat q = \sqrt{\frac{\hbar}{2 M \nu}}\left(\hat a + \hat a^\dagger\right)$, where $M$ is the mass of the ion, and $\mu=\pm 1$ is the sign of the detuning.

Moving to an interaction picture with respect to both qubits and the motional mode and dropping terms that rotate at twice the qubit frequency (rotating wave approximation), the Hamiltonian can be written as
\begin{equation}
    \label{eq:Hamiltonian normal}
    \hat H = \hbar\Omega(t) \cos(\delta t)\left(\hat J_+ e^{ik\hat q(t)} + \hat J_- e^{-ik\hat q(t)}\right),
\end{equation}
where $\hat J_{\pm} = \left(\hat J_x \pm i\hat J_y\right)$ are the collective spin raising and lowering operators.

In the following sections, we need to calculate integrals of products of the Hamiltonian at different times. Thus, it is useful to expand the Hamiltonian as a power series in the raising and lowering operators that act on the motional mode subspaces. 
To achieve this, we first express the operator for the motional mode in terms of the creation and annihilation operators ($\hat a^\dagger$, $\hat a$) as $k\hat q(t) = \eta\left(\hat a e^{-i\nu t}+\hat a^\dagger e^{i\nu t}\right)$, where $\eta$ is the Lamb-Dicke factor, which describes the strength of the coupling between the electronic state and the motional mode of the ion.
Then, we use the Baker-Campbell-Hausdorff (BCH) formula \cite{hall_lie_2015} to factorize the operator exponentials as
\begin{equation}
    \label{eq:BCH trap}
    e^{ i\eta\left(\hat a e^{-i \nu t} + \hat a^\dagger e^{i\nu t}\right)} = e^{ i \eta \hat a^\dagger e^{i\nu t}} e^{i \eta \hat a e^{-i\nu t}}e^{ \frac{-\eta^2}{2}}.
\end{equation}
Next, we expand both exponentials in \cref{eq:BCH trap} via Taylor series and group terms together with an index $m$ for the sideband order (i.e.~the number of phonons created or annihilated). 
Meanwhile, we Fourier-decompose the time-dependent parts in \cref{eq:Hamiltonian normal} and write the drive amplitude over the time interval $[0,T]$ as a Fourier series
\begin{equation} \label{eq:drive amplitude Fourier}
    \Omega(t) = \Omega \sum_{M\in \mathbb Z} c_M e^{i\frac{2\pi}{T}Mt},
\end{equation}
with $c_{-M}=c_M^*$, and where in principle we truncate the series at a frequency for bandwidth considerations.
Inserting into \cref{eq:Hamiltonian normal}, this leads to
\begin{equation}
    \label{eq:Hamiltonian factored}
	\hat H(t) = \hbar\Omega \sum_{M\in\mathbb{Z}} \sum_{m\in\mathbb{Z}}\sum_{\mu=\pm 1} c_M e^{i\left(\frac{2\pi}{T}M+m\nu+\mu \delta\right)t} \hat J_m\hat A_m,
\end{equation}
where operations on the qubit subspace are described by the collective spin operator,
\begin{equation}
		\hat J_m = \frac{1}{2}\left(\hat J_+ + (-1)^m \hat J_-\right) =\begin{cases}
          \hat J_x &\quad \text{if } m \text { even}\\
          i\hat J_y &\quad \text{if } m \text { odd}
      \end{cases}, \label{eq:Jm}
\end{equation}
 which acts on both qubits. The transition in the motional mode under absorption or emission of $m$ phonons is given by the $m$-th sideband  transition operator 
\begin{equation}
     \hat A_m = e^{-\frac{1}{2}\eta^{2}}\sum_{k=\max(0,-m)}^\infty\frac{\eta^{2k+m}i^{2k+m}}{(m+k)!k!}(\hat{a}^{\dagger})^{k+m}(\hat{a})^{k}.\label{eq:Am}
 \end{equation}

We next proceed to calculate the Magnus expansion for the Hamiltonian \cref{eq:Hamiltonian factored}, order by order, so that we can see the effect of each term, and how some terms can be corrected.

\section{Resonance conditions}\label{sec:Resonance Conditions}
Our goal is to set conditions to suppress certain transitions in \cref{fig:Energy level diagram}, while enabling the required MSG qubit double-excitation transitions. For this purpose, we associate the resonance conditions with beat-note frequencies
\begin{equation}\label{eq:Nj}
    N(m,\mu)= \underbrace{\frac{\nu T}{2\pi}}_K m + \underbrace{\frac{\delta T}{2\pi}}_L\mu,
\end{equation}
with dimensionless gate duration $K$ and dimensionless laser detuning frequency $L$. As before, $m$ is the sideband order and $\mu$ is the detuning sign. Only the zeroth harmonic of the pulse is included in this section, i.e.~$M=0$, for better readability. The generalized versions of the formulas are provided in \cref{sec:Resonance integral shaped}.

In this notation, the Hamiltonian \cref{eq:Hamiltonian factored} takes the simplified form
\begin{equation}
    \label{eq:Hamilton with integer}
	\hat H(t) = \hbar\Omega \sum_{m=-\infty}^\infty\sum_{\mu=\pm1}   e^{i\frac{2\pi}{T} N(m,\mu) t}\hat J_m\hat A_m.
\end{equation}

 While the Magnus expansion can be very cumbersome beyond second order, we benefit here from the simplification that $Z_1=0$, which is required to remove single qubit driving, whose details are given below. In effect, this means that, using \cref{eq:Magnus terms for P1=0}, calculation of the $k$-th order Magnus expansion terms can be reduced in its time dependence to solving nested integrals of the form 
\begin{equation}\label{eq:resonance integral}
    I_{N_1,\cdots,N_k} =\int_0^T \dd t_k \int_0^{t_k} \dd t_{k-1} \cdots \int_0^{t_2} \dd t_1 e^{i\frac{2\pi}{T}\sum_{j=1}^k N_j t_j},
\end{equation}
where the exponent consists of the sum of the exponents of \cref{eq:Hamilton with integer} evaluated at different times $t_j$ and for different combinations of the beat-note frequencies $N_j  \equiv N(m_j,\mu_j)$. Finding appropriate solutions to the conditions will allow us then to significantly simplify the time-ordered operators in \cref{eq:Dyson terms}, rewritten here as
\begin{equation}\label{eq:dyson term factored}
    \hat P_k(t)=\left(-\frac{i}{\hbar}\right)^k\prod_{j=1}^{k}\bigg(\sum_{m_j=-\infty}^\infty\sum_{\mu_j=\pm1} \hat J_{m_j}\hat A_{m_j} I_{N_j} \bigg),
\end{equation}
which factors out the time dependence.  

We proceed to calculate the leading orders of the nested integrals analytically and then verify their accuracy numerically. A main challenge, especially for fast oscillating numerics, is to ascertain whether for single coefficients $N_j$ the integrals are zero, and whether they combine to sum up to zero, each corresponding to different resonance conditions.

For the evaluation of the higher order terms of the Magnus expansion in this paper, we thus used integration by parts to simplify the integrals.
The analytical forms and their numerical evaluation can then be calculated using the SymPy \cite{meurer_sympy_2017} and NumPy \cite{harris_array_2020} python packages, respectively.

\subsection{First order condition}
The resonance condition for the first Magnus order results from the integral $I_{N_1}$, which produces $\hat Z_1=i\hat P_1$ given by 
\begin{equation}
    \label{eq:Z1}
	\hat Z_1 = \Omega\sum_{m=-\infty}^\infty\sum_{\mu=\pm1} \underbrace{\frac{e^{i2\pi N(m, \mu)}-1}{i\left(m\nu+\mu \delta\right)}}_{\text{resonance condition}}\hat J_m\hat A_m.
\end{equation}
This term describes the phonon-assisted flipping of the qubits. For the entangling gate, we want to suppress single qubit flips at the final time $T$, so we tune the parameters such that $\hat Z_1 = 0$. This requires that the beat-note index must be a nonzero integer
\begin{equation}\label{eq:firstordercondition}
    N(m,\mu) \in \mathbb{Z}^*,
\end{equation}
which implies that also $K$ and $L$ are integers.
For constant drive amplitude $M=0$, fulfilling \cref{eq:firstordercondition} can be ensured with a laser detuning $\delta$ that is not a multiple of the trap frequency $\nu$. 

Note that due to strong driving (compared to the detuning), the motional mode will nonetheless be excited, and only at the final time does it revert to zero.
If, on the other hand, the laser detuning is on resonance with the motional mode frequency, $\delta=\nu$, the sideband transition remains populated also at the final time \cite{wineland_experimental_1998}. Small detuning of the laser frequency will violate \cref{eq:firstordercondition} and thus lead to errors in $\hat Z_1$ \cite{martinez-garcia_analytical_2022}, but is outside the scope of the current work.

\begin{figure}[t]
    \centering
    \includegraphics[width=1\linewidth]{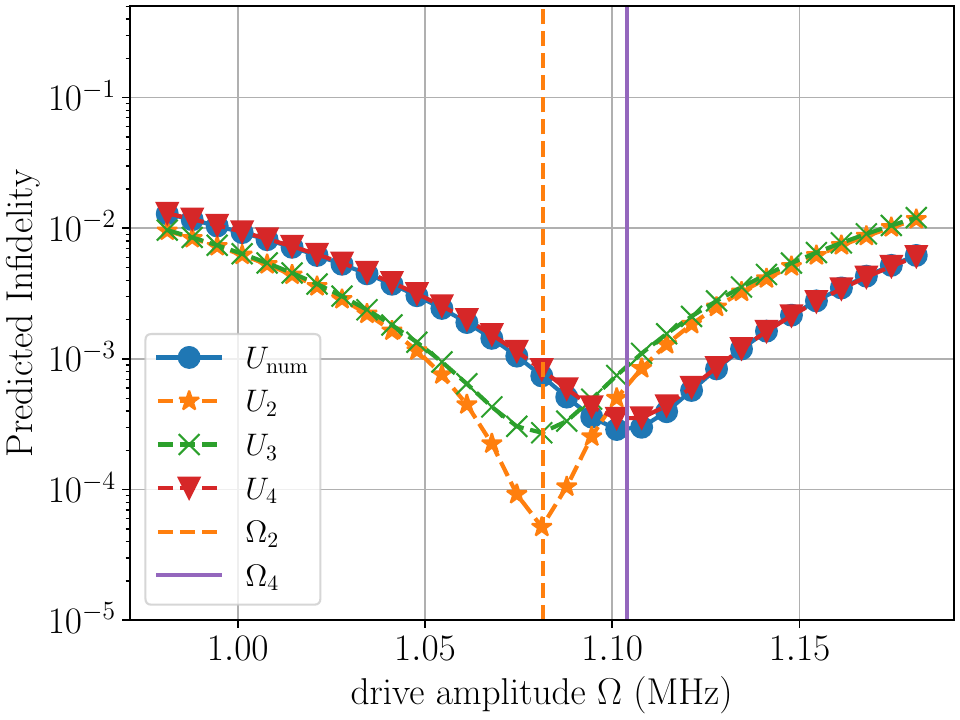}
	\caption{Predicted average infidelity as a function of the drive amplitude $\Omega$ calculated with $\hat U_\text{num}$ and different Magnus expansion orders $\hat U_2,\hat U_3,\hat U_4$ in \cref{eq:def Un}. The parameters are $\eta=0.18$, $K-L=\frac{\nu-\delta}{2\pi}T$=3, $K=28$, $\nu/(2\pi) = \SI{1}{MHz}$, and $\bar n=\num{2e-2}$. The theoretical prediction of $\Omega_2$ (\cref{eq:Omega opt U2}) is depicted with the dashed orange vertical line. The theoretical prediction of the minimum $\Omega_4$ (\cref{eq:Omega opt U4}) is shown with the solid purple vertical line. 
    The Magnus orders $\hat Z_k$ are calculated taking into account the first three sideband orders $|m_i|\leq3$. }
    \label{fig:Errors rect Omega}
\end{figure}

\subsection{Second order condition}
In the second order, we plug the Dyson term, \cref{eq:dyson term factored}, into the Magnus expansion, \cref{eq:Magnus}, and obtain

\begin{equation}\label{eq:Z2 general}\begin{aligned}
\hat Z_2 =  \frac{i}{2}\Omega^2\sum_{m_1, m_2=-\infty}^\infty\sum_{\mu_1, \mu_2=\pm 1}\
\hat J_{m_1}\hat J_{m_2} \hat A_{m_1}\hat A_{m_2}I_{N_1,N_2},
\end{aligned}
\end{equation}
with the resonance integral $I_{N_1,N_2}$  given by \cref{eq:resonance integral}.

We focus on the case where $N_j\in \mathbb Z,\forall j$. Then, the non-vanishing contributions to the integral are:
\begin{equation}\label{eq:Int 2}
   I_{N_1,N_2}  = \begin{cases}
        \frac{T^2}{2}&\quad N_1=0,N_2=0\\
        \frac{iT^2}{2\pi N_2}&\quad N_1=0,N_2\neq0\\
        -\frac{iT^2}{2\pi N_1}&\quad N_1\neq0,N_2=0\\
        -\frac{iT^2}{2\pi N_2}&\quad N_1\neq 0,N_2\neq0,N_1+N_2=0.
    \end{cases}
\end{equation}
Here, we have four resonance conditions. $N_j=0$ are the resonance conditions for the unwanted single-photon transitions.
The fourth condition, where $N_j\neq 0$ and $N_1+N_2=0$, is the resonance condition for a two-photon transition. This is what creates the desired  \moso{} entanglement.

The expression for the second order of the Magnus expansion, \cref{eq:Z2 general}, can be further simplified if the single photon transitions, \cref{eq:Z1}, are suppressed.
Assuming that the detuning $\delta$ of the laser from the qubit frequency is of the order of the trap frequency $\nu$, which is required anyway in order to excite a single motional mode only, then the two-photon resonance condition can be rewritten as
\begin{equation}
\label{eq:multipartite resonance condition}
\begin{split}
    N_1+N_2=0\iff  (m_1=-m_2) \land (\mu_1=-\mu_2).
\end{split}
\end{equation}
In \cref{sec:Magnus term condition}, we discuss this matter in more detail.
The simplification in \cref{eq:multipartite resonance condition} allows rewriting of \cref{eq:Z2 general} as
\begin{equation}
    \label{eq:Z2}
    \hat Z_2 =  \sum_n \left(\hat J_x^2 d_x^{(n)} + \hat J_y^2 d_y^{(n)} \right)\otimes \dyad{n}{n},
\end{equation}
with the standard form factors  \cite{sorensen_entanglement_2000} that are the sum over even $m$ for $d_x$ and over odd $m$ for $d_y$,
\begin{eqnarray}
    \label{eq:Z2 coeffs}
    d_{x,y}^{(n)} = \mp\frac{\Omega^2T^2}{2\pi} e^{-\eta^2} &\sum_{m \text{ even, odd}}\sum_{\mu=\pm1} \frac{1 }{mK+\mu L} (-\eta^2)^{|m|} \nonumber\\
    & \left(L_{\min(n,n-m)}^{|m|}(\eta^2) \right)^2\frac{\min(n,n-m)!}{\max(n,n-m)!},\nonumber\\
\end{eqnarray}
where $L_a^{(b)}$ is the associated Laguerre polynomial \cite{wineland_experimental_1998}.

Note that \cref{eq:Z2} is diagonal in the motional mode subspace.
Therefore, as can be seen from \cref{fig:Energy level diagram}, the terms in $\hat Z_2^{(m=1)}$ only perform flips (simultaneously) on the two qubits. 

However, these conditions alone do not ensure that the Hamiltonian generates the correct gate. As we discuss next in Sec.~\ref{sec:Error terms}, recalibrating the weights of these and other terms affects the rotation angle about the $\hat J_x^2$ and $\hat J_y^2$ operators.

\subsection {Higher orders}
When $\hat Z_1=0$, the third order of the Magnus expansion similarly simplifies to $\hat Z_3 = i\hat P_3$. The calculation of this term involves calculation of a triple integral. Assuming that we suppress single-photon transitions, $N_j\in  \mathbb Z^*\;\forall j$, the resonance integral of third order is
\begin{equation}
\label{eq:Int 3}
\begin{split}
I_{N_1,N_2,N_3}
= \frac{T^3}{4\pi^2 N_3}\left(\frac{\delta_{N_1+N_2}}{N_2}+\frac{\delta_{N_2+N_3}}{N_1}\right).
\end{split}
\end{equation}
This integral is zero unless $N_1+N_2=0$ or $N_2+N_3=0$, which are the resonance conditions for the transitions.
Because of the complexity of \cref{eq:Int 3}, there is no compact form for $\hat Z_3$ beyond \cref{eq:dyson term factored}, as there was for $\hat Z_2$ in \cref{eq:Z2}. The same applies for $\hat Z_4$ and $\hat Z_5$, but we will provide the leading analytical error terms for $\hat Z_3$ and $\hat Z_4$, for the typical case of a rectangular pulse next, in \cref{sec:analytical results}. We then evaluate numerically the significance of those terms in \cref{sec:Numerical Results rect}.

\section{Error terms for rectangular drive pulses}\label{sec:Error terms}

In this section, we present the leading order terms of the Magnus expansion and Lamb-Dicke parameter $\eta$ for the \moso{} driven with rectangular pulses. We examine the impact and correction possibilities in terms of the average fidelity.

\begin{table*}[ht]
    \centering
    \renewcommand{\arraystretch}{2.5} 
    \setlength\tabcolsep{2pt}
    \begin{tabular}{cccccc}
        Error & Operator & Term & Term at $\Omega_{\text{LD}}$ & Term at $\Omega_{4}$ 
		\tabularnewline
        \hline 
        \hline 
        Gate & $\hat J_{y}^{2}$ & 
        $-\frac{K\Omega^{2}T^{2}\eta^{2}}{\pi\left(K^{2}-L^{2}\right)}$ & 
        $-\frac{\pi}{2}$ & 
        $-\frac{\sqrt{2K} \pi L \eta\left(s - \sqrt{ s^{2}- (K^{2} -L^{2} )}\right)}{\left(K^2 - L^2\right)}$ &
		\tabularnewline
        \hline 
        $\hat Z_{2}^{(m=1)}$ &
        $\hat J_{y}^{2}$ & 
        $\frac{K\Omega^{2}T^{2}\eta^{4}(2n+1)}{\pi\left(K^{2}-L^{2}\right)}$ &
        $\frac{\pi\eta^{2}(1+2n)}{2}$ & 
        $\frac{\sqrt{2K} \pi L \eta^3 \left(2n+ 1\right) \left(s - \sqrt{ s^{2}- (K^{2} -L^{2} )}\right)}{\left(K^2 - L^2\right)}$&
		\tabularnewline
        \hline 
        $\hat{Z}_{2}^{(m=2)}$ &
        $\hat J_{x}^{2}$ & 
        $-\frac{K\Omega^{2}T^{2}\eta^{4}(2n+1)}{\pi\left(4K^{2}-L^{2}\right)}$ &
        $-\frac{\pi\eta^{2}\left(K^{2}-L^{2}\right)(2n+1)}{2 \left(4K^{2}-L^{2}\right)}$ & 
        $-\frac{\sqrt{2K} \pi L \eta^{3} (2n+1) \left(s - \sqrt{ s^{2}- (K^{2} - L^{2})}\right)}{4 K^{2} - L^{2}}$&
		\tabularnewline
        \hline 
        $\hat{Z}_{3}^{(m=1)}$  &
		$\hat J_{y}(\hat a+\hat a^{\dagger})$ & $-\frac{2K^{2}\Omega^{3}T^{3}\eta^{5}}{\pi^{2}\left(K^{2}-L^{2}\right)^{2}}$ & $-\pi\eta^2\sqrt{\frac{K}{2\left(K^2-L^2\right)}}$ & $- \frac{2^{\frac{7}{4}} \pi K^{\frac{5}{4}} L^{\frac{3}{2}} \eta^{\frac{7}{2}} \left(s - \sqrt{- K^{2} + L^{2} + s^{2}}\right)^{\frac{3}{2}}}{(K^{2} -  L^{2} )^2}$&
		\tabularnewline
        \hline 
        $\hat{Z}_{3}^{(m=2)}$   &
		$\hat J_x\left(\mathds{1}-\hat J_y^2\right)(\hat a^{2}-\hat a^{\dagger2})$ & $\frac{K^{2}\Omega^{3}T^{3}\eta^{4}}{\pi^{2} \left(4K^{2}-L^{2}\right)\left(K^{2}-L^{2}\right)}$ & $\frac{\pi\eta}{2}\sqrt{\frac{K}{2\left(K^2-L^2\right)\left(4K^2-L^2\right)}}$ & 
        $-\frac{2^{\frac{3}{4}} \pi K^{\frac{5}{4}} L^{\frac{3}{2}} \eta^{\frac{5}{2}} \left(s - \sqrt{s^{2}- (K^{2} - L^{2} ) }\right)^{\frac{3}{2}}}{(4 K^{2} -  L^{2})(K^2-L^2)}$&
		\tabularnewline
        \hline 
        \multirow{3}{*}{$\hat{Z}_{4}^{(m=1)}$ } & 
        $\hat J_{xy}(\hat a+\hat a^{\dagger})$ & 
        $\frac{K\Omega^{4}T^{4}\eta^{3}}{\pi^{3}\left(K^{2}-4L^{2}\right)\left(K^{2}-L^{2}\right)}$ & 
        $\frac{\pi\left(K^{2}-L^{2}\right)}{4K\eta\left(K^{2}-4L^{2}\right)}$ & 
        $\frac{2 \pi L^{2} \eta \left(s - \sqrt{ s^{2}- (K^{2} - L^{2} )}\right)^{2}}{(K^{2} -L^{2} )(K^2- 4 L^{4})}$&
		\tabularnewline
        \cline{2-6} 
         & $\hat J_{z}^{2}$ & 
         $-\frac{K\Omega^{4}T^{4}\eta^{2}}{4\pi^{3}L^2\left(K^{2}-4L^{2}\right)}$ & 
         $-\frac{\pi\left(K^{2}-L^{2}\right)^2}{16KL^2\eta^{2}\left(K^{2}-4L^{2}\right)}$ & 
         $\frac{\pi \left(K^{2} - L^{2} - 2 s^{2} + 2 s \sqrt{ s^{2}- (K^{2} - L^{2})}\right)}{2 \left(K^{2} - 4 L^{2}\right)}$&
		 \tabularnewline
        \cline{2-6} 
         & $\hat J_{y}^{2}$ & 
         $\frac{K\Omega^{4}T^{4}\eta^{2}}{4\pi^{3}L^{2}\left(K^{2}-L^{2}\right)}$ & 
         $\frac{\pi\left(K^{2}-L^{2}\right)}{16KL^{2}\eta^{2}}$ & 
         $-\frac{\pi}{2} + \frac{\pi  s \left(s -  \sqrt{ s^{2}- (K^{2} - L^{2})}\right)}{ \left(K^{2} - L^{2}\right)}$&
		 \tabularnewline
    \end{tabular}
    \caption{Error terms for rectangular pulses. The second column shows the operator part of the term. The 3rd to 5th column show the coefficient of the operator. The 4th and 5th column are the same as the 3rd, except that $\Omega=\Omega_{\text{LD}}$ (\cref{eq:Omega opt LD}), and $\Omega=\Omega_4$ (\cref{eq:Omega4} with $s = \sqrt{2K}L\eta(1-\eta^2)$), respectively. 
}
    \label{tab:Error budget}
\end{table*}

\subsection{Analytical Expressions}\label{sec:analytical results}
The analytical expressions for the error terms are calculated by solving the time-ordered Dyson integrals \cref{eq:dyson term factored} for small sideband numbers $m$. 
Moreover, when neglecting terms of the Magnus expansion that are off-diagonal in the motional mode subspace, we can also derive analytical expressions for the fidelity, as explained in detail in \cref{sec:choice of fidelity}.

When considering only terms in the Hamiltonian which are proportional to $\hat J_y^2$, a simple expression for the average fidelity can be obtained, see \cref{eq:average fidelity dy}.
By employing the Bell fidelity, a simple expression can be derived for the case that terms proportional to $\hat J_x^2$ and $\hat J_z$ are taken into account as well, see \cref{eq:Bell fidelity sin dx dy}.
Both expressions yield simple equations that can be solved for $\Omega$ to determine the value that optimizes the fidelity.

\subsubsection{Errors and optimal drive amplitude at $\hat Z_2$}

To the first nonzero order of $\eta$, taking $k=0$ and the first sideband order $|m|= 1$ in \cref{eq:Am}, the form factors in \cref{eq:Z2} simplify, such that  $d_x^{(0)}=0$ and the second order Magnus term is $\hat Z_2= d_y^{(0)}\hat J_y^2=-\frac{\Omega^2T^2}{\pi}\frac{K}{(K^2-L^2)}\eta^2\hat J_y^2$. 

The optimal parameters that fulfill a $\pi/2$ rotation around $\hat J_y^2$ can straightforwardly be solved (c.f.~\cref{sec:choice of fidelity}), leaving us with

\begin{equation}
    \label{eq:Omega opt LD}
    \Omega_\text{LD}= \frac{\pi}{ T \eta}\frac{ \sqrt{(K^{2} - L^{2})}}{\sqrt{2K}}, 
\end{equation}
where we assume that the motional mode is initially in the vacuum state $\bar n=0$.

Taking the next order of the Lamb-Dicke expansion into account gives the Lamb-Dicke error term 
\begin{equation}\label{eq:Z2 m1}
    \hat Z_2^{(m=1)}=\frac{K\Omega^{2}T^{2}\eta^{4}(2n+1)}{\pi\left(K^{2}-L^{2}\right)}\hat J_y^2 = \hat J_y^2 \bigO{\Omega^2\eta^4},
\end{equation}
which corresponds to the 2nd row in \cref{tab:Error budget}.
Taking also the next sideband order into account, with $|m|=2$ in \cref{eq:Z2 coeffs}, gives the sideband error term (3rd row in \cref{tab:Error budget})
\begin{equation}\label{eq:Z2 m2}
    \hat Z_2^{(m=2)}= -\frac{K\Omega^{2}T^{2}\eta^{4}}{\pi\left(4K^{2}-L^{2}\right)}(2n+1)\hat J_x^2 =\hat J_x^2 \bigO{\Omega^2\eta^4}.
\end{equation}
Both errors have already been studied in the original work, Ref.~\cite{sorensen_entanglement_2000}, and can be minimized by further adjusting the drive amplitude
\begin{equation}\label{eq:Omega opt U2}
    \Omega_{2} = \frac{ \pi }{ T \eta}\sqrt{\frac{(K^2-L^2)(4K^2-L^2)}{ 2K\left( \eta^{2} (2 L^{2}-5 K^{2})+ ( 4 K^{2} - L^{2})\right)}},
\end{equation}
which can be found by inserting the form factors in \cref{eq:Z2} for $n=0$ into the condition for the optimal fidelity (see \cref{sec:choice of fidelity} for more details). 

However, these errors are only optimal in the approximation of zero motional mode population $\bar n=0$, and discounting the next orders of the Lamb-Dicke and Magnus expansion. We now discuss these latter corrections.

\begin{figure*}[ht]
    \centering
    \begin{subfigure}[h]{0.32\textwidth}
        \centering
        \includegraphics[width=1\linewidth]{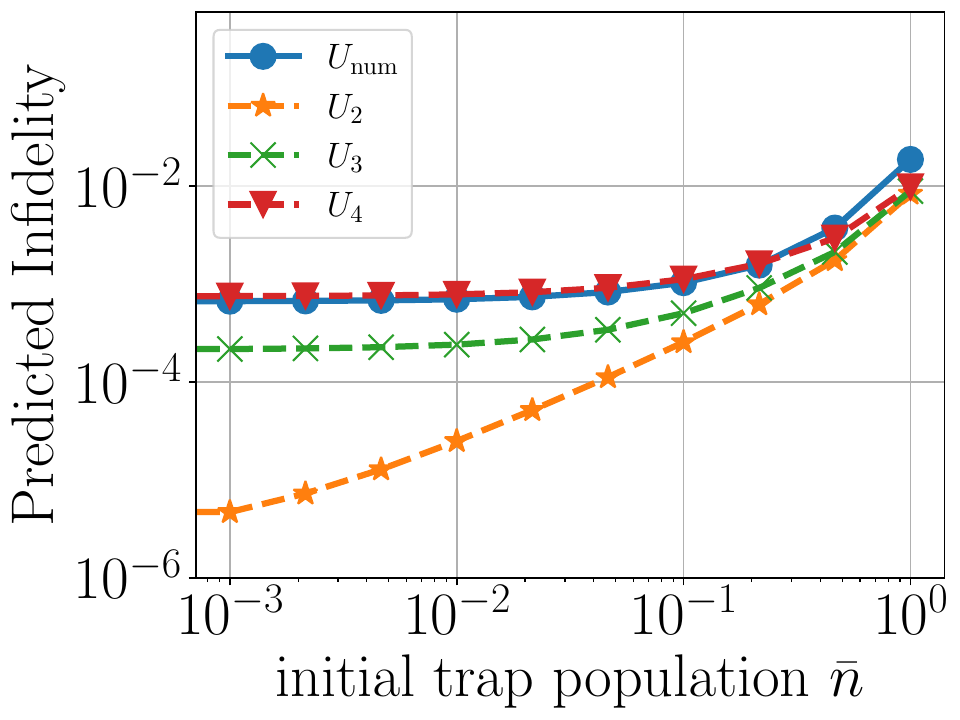}
        \caption{}
        \label{fig:Errors rect nbar}
    \end{subfigure}
    \begin{subfigure}[h]{0.32\textwidth}
        \centering
        \includegraphics[width=1\linewidth]{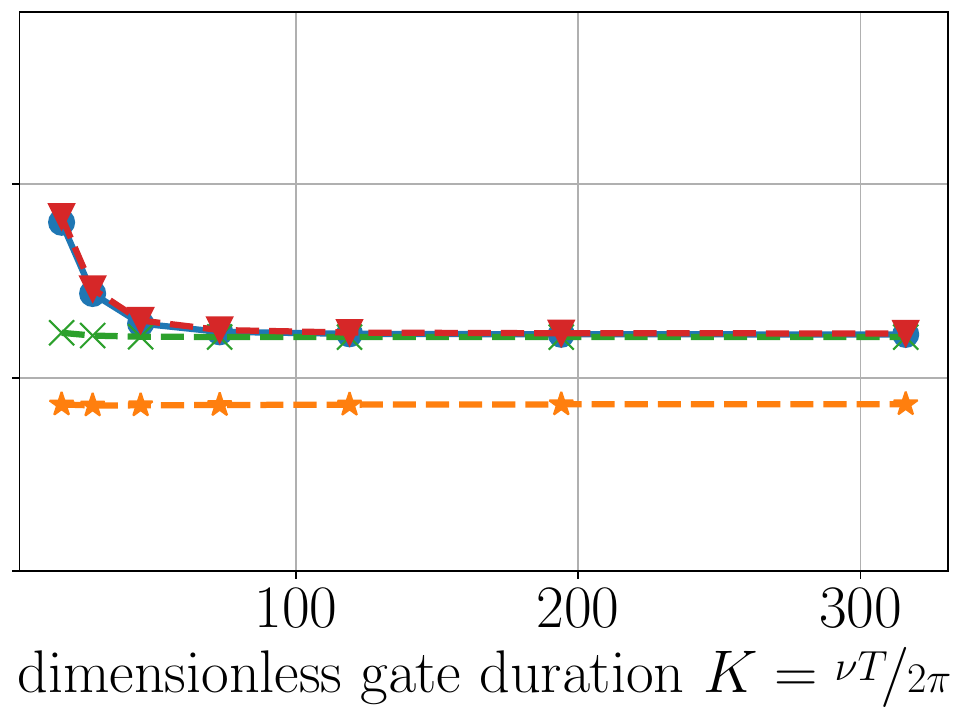}
       \caption{}
        \label{fig:Errors rect K}
    \end{subfigure}
    \begin{subfigure}[h]{0.32\textwidth}
        \centering
        \includegraphics[width=1\linewidth]{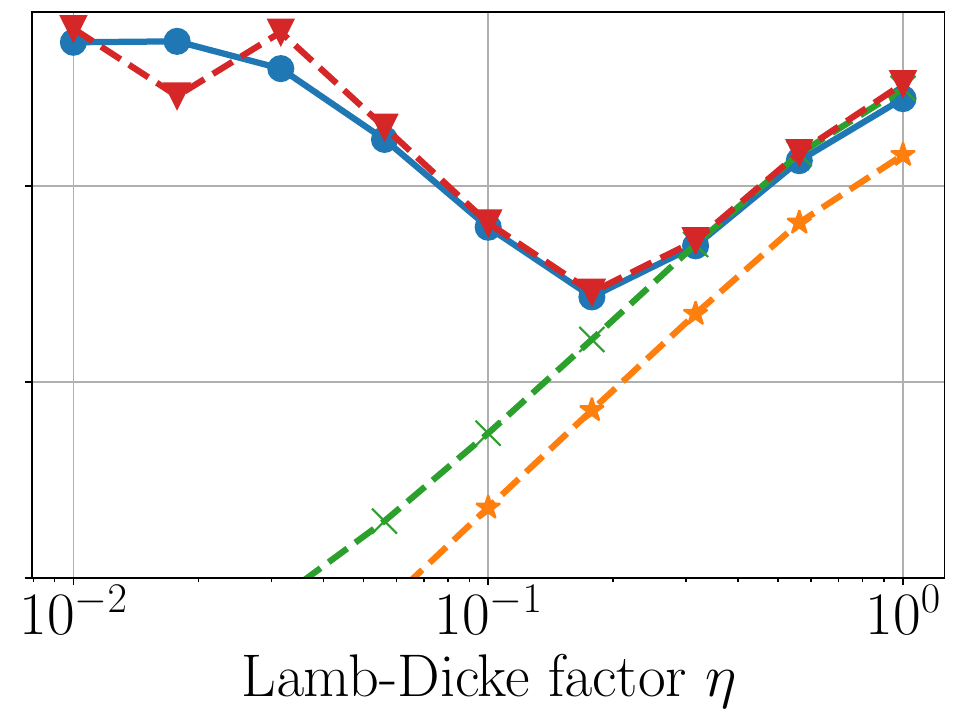}
      \caption{}
        \label{fig:Errors rect eta}
    \end{subfigure}
	\caption{Predicted average infidelity as a function of the population of the motional mode $\bar n$ (\cref{fig:Errors rect nbar}), the dimensionless gate duration $K$ (\cref{fig:Errors rect K}), and the coupling strength $\eta$ (\cref{fig:Errors rect eta}), respectively. The parameters are (if they are not varied): $\eta=0.18$, $K-L=\frac{\nu-\delta}{2\pi}T=3$, $K=28$, $\nu/(2\pi)=\SI{1}{\mega\hertz}$, $\bar n=\num{2e-2}$, and $\Omega=\Omega_2$ (\cref{eq:Omega opt U2}). $U_n$ and $U_\text{num}$ are the predicted infidelities calculated with \cref{eq:def Un} and \cref{eq:def Unum}, respectively.
    }
    \label{fig:Errors rect}
\end{figure*}

\subsubsection{Errors at $\hat Z_3$}
Now let us take a look at the higher orders of the Magnus expansion. In this subsection and the following, we discuss the error operators and the order of the terms in $\Omega$ and $\eta$, while full expressions are shown in \cref{tab:Error budget} for better readability.

The first-sideband term of $\hat Z_3$ (4th row in \cref{tab:Error budget}) for a rectangular pulse takes the form
\begin{equation}\label{eq:Z3 m1}
    \hat Z_3^{(m=1)} = \hat J_y(\hat a+\hat a^\dagger) \bigO{\Omega^3\eta^5},
\end{equation}
which gives $\bigO{\eta^2}$ at $\Omega=\Omega_\text{LD}$. This term creates a phonon-assisted single-qubit flip. 
The second-sideband term for $\hat Z_3$ (5th row in \cref{tab:Error budget}) is of order
\begin{equation}\label{eq:Z3 m2}
		\hat Z_3^{(m=2)} = \hat J_{x}\left(\mathds{1}-\hat J_y^2\right)\left(\hat a^2-(\hat a^\dagger)^2\right) \bigO{\Omega^3\eta^4}.
\end{equation}
which gives $\bigO{\eta}$ at $\Omega=\Omega_\text{LD}$. This term causes squeezing \cite{katz_body_2022}, and population leakage into the motional mode.
The energy level diagram in \cref{fig:Energy level diagram} depicts how the processes in $\hat Z_3$ cause population leakage in both the motional mode and qubit subspaces.

\subsubsection{Errors and optimal drive amplitude at $\hat Z_4$}
The fourth order of the Magnus expansion consists in the sum of three entangling operations
\begin{equation}\label{eq:Z4 m1}
\begin{aligned}
    \hat Z_4^{(m=1)} = &\hat J_{xy}\left(\hat a + \hat a^{\dagger}\right) \bigO{\Omega^4\eta^3} \\&+
   \hat J_{z}^{2} \bigO{\Omega^4\eta^2} \\&+ \hat J_{y}^{2}  \bigO{\Omega^4\eta^2},
\end{aligned}
\end{equation}
where 

    $\hat J_{\alpha \beta} = \frac{1}{2}\left(\hat \sigma_\alpha\otimes \hat \sigma_\beta + \hat \sigma_\beta \otimes \hat \sigma_\alpha\right).$

The full expressions are listed in rows 6–8 of \cref{tab:Error budget}.
As sketched in \cref{fig:Energy level diagram}, the first term in $\hat Z_4$, $\hat J_{xy}\left(\hat a + \hat a^\dagger\right)$, creates entanglement, but also causes population leakage in the motional mode. The second term, $\hat J_z^2$, only adds a phase to the target state. The third term, $\hat J_y^2$, contributes to the entangling gate.
The leading terms of the first sideband order terms $\hat Z_2^{(m=1)}+\hat Z_3^{(m=1)} + \hat Z_4^{(m=1)}$ are proportional to $\hat J_y^2$.
Inserting the resulting prefactor
\begin{equation}
		\begin{aligned}
				d_y^{(n)} &= -\Omega^2T^2\frac{K\eta^2(\eta^2-1)}{\pi\left(K^2-L^2\right)} + \Omega^4T^4 \frac{K\eta^2}{4\pi^3L^2\left(K^2-L^2\right)}
		\end{aligned}
\end{equation}
into the condition for the Bell fidelity \cref{eq:Bell fidelity sin dx dy} or for the average fidelity \cref{eq:average fidelity dy} and setting $\mathcal{F} = 1$ yields a quadratic equation in $\Omega^2$,
\begin{equation}
- \frac{K \Omega^{2} T^{2} \eta^{2} \left(4 \pi^{2} L^{2} \eta^{2} - 4 \pi^{2} L^{2} + \Omega^{2} T^{2}\right)}{4 \pi^{3} L^{2} \left(K^{2} - L^{2}\right)} = \frac{\pi}{2},\label{eq:Omega opt U4}
\end{equation}
whose solution is the corrected optimal drive amplitude 
\begin{equation}
    \Omega_4^2 = \frac{\sqrt{2}\pi^2L\left(s-\sqrt{s^2-(K^2-L^2)} \right)}{\sqrt{K}T^2\eta},
    \label{eq:Omega4}
\end{equation}
with $s = \sqrt{2K}L\eta(1-\eta^2)$.

Since $\Omega_4$ is $\bigO{\eta^{-{\frac{1}{2}}}}$, in contrast to $\Omega_\text{LD}$ and $\Omega_2$ which are $\bigO{\eta^{-1}}$, the Lamb-Dicke orders of the terms of the Magnus expansion change when driving at $\Omega_4$ instead of $\Omega_2$ or $\Omega_\text{LD}$, which is reflected in the last two columns of \cref{tab:Error budget}.

\subsection{Numerical Comparison}\label{sec:Numerical Results rect}

In the previous section, the analytical expressions for the leading error terms in the Magnus expansion were presented, summarized in \cref{tab:Error budget}.
This section focuses on numerically analyzing the impact of these error terms on the average fidelity (\cref{eq:weighted average fidelity}) of the \moso{}.
We compare the infidelities $1-\mathcal{F}_\text{av}$ calculated with different propagators.
The fidelity representing the effect of the $n$-th order of the Magnus expansion is calculated with
\begin{equation}\label{eq:def Un}
    \hat U_n := \exp\left\{-\frac{i}{\hbar}\sum_{k=2}^n \hat Z_k\right\}.
\end{equation}
The first order $\hat Z_1$ is always suppressed.
The numerical expression for the Magnus terms $\hat Z_n$ are calculated using the formulas shown in the previous sections.

The operators are implemented in matrix form, with the operators acting on the motional mode subspace being truncated at $N_\text{dim}=8$.
The resonance integrals are calculated recursively using computer tools as described in Section~\ref{sec:Resonance Conditions}.

In the previous section, the analysis of the Magnus expansion orders was limited to the second to fourth order. 
To ensure the adequacy of this restriction, we calculated the infidelity including the fifth-order Magnus term as well, represented as $\hat U_5$.
However, since the results show that the curves obtained with $\hat U_5$ nearly completely overlap with those calculated with $\hat U_4$, we have decided to not include them into the figures in order to avoid visual clutter.

To evaluate the accuracy of a truncated propagator in representing the actual system evolution, we estimate the propagator for infinite Magnus orders $\hat U_\infty$. This estimation is accomplished through numerical calculation of the propagator using the Trotter expansion \cite{suzuki_general_1992,trotter_product_1959} which decomposes the propagator into products of exponential operators, each corresponding to a small time step. We define the numerical propagator as
\begin{equation}\label{eq:def Unum}
   \hat U_\text{num} := \prod_{n=0}^{N_t} e^{-\frac{i}{\hbar} H(n\Delta t)\Delta t}.
\end{equation}
To ensure that the time steps are sufficiently small, they are set to $\Delta t = \frac{1}{10 f_\text{max}}$ where $f_\text{max} = \max \{M + mK + L\} \frac{2\pi}{T}$ is the largest frequency appearing in the Hamiltonian, \cref{eq:Hamilton with integer}. As the Magnus expansion terms for the numerical calculations are truncated after the third sideband order, we set $m=3$ here.

The figures \cref{fig:Errors rect Omega,fig:Errors rect,fig:Errors sin2} were created using the following typical parameters \cite{mehta_integrated_2020,zarantonello_robust_2019,wong-campos_demonstration_2017,schafer_fast_2018}:
The Lamb-Dicke parameter is set to $\eta=0.18$. 
The average trap population is chosen as $\bar n = \num{2e-2}$. 
For rectangular pulses, the dimensionless gate durations is $K=28$, which corresponds to a gate duration of \SI{28}{\micro\second} for a trap frequency of $\nu/(2\pi)=\SI{1}{\mega\hertz}$.
The dimensionless laser detuning difference from the trap frequency $K-L$ is set to 3, corresponding to a detuning of $(\nu-\delta)/(2\pi) = \SI{107}{\kilo\hertz}$, illustrating that the smallest detuning of $K-L=1$ may not always be optimal. 
Because the difference in fidelity for different values of $K-L$ is not significant and its behavior depends on the choice of other parameters, this article does not delve into the specifics of how the choice of $K-L$ affects the fidelity.

In \cref{fig:Errors rect Omega}, the predicted infidelities for the propagators $\hat U_n$ and $\hat U_\text{num}$ are displayed as a function of the drive amplitude $\Omega$. The minima of the predicted infidelities calculated with $\hat U_2$ and $\hat U_3$, respectively, are approximately located at the drive amplitude $\Omega_2$, whose value was calculated using \cref{eq:Omega opt U2}.
 The curves for $\hat U_4$ and $\hat U_\text{num}$ are nearly indistinguishable at off-resonance and are closely aligned around their minimum, which is approximately at the expected optimal Rabi frequency $\Omega_4$ (\cref{eq:Omega opt U4}.
The predicted infidelity of $\hat U_3$ at $\Omega_2$ is approximately equal to the infidelities of $\hat U_4$ and $\hat U_\text{num}$ at $\Omega_4$.
For the sake of consistency, in \cref{fig:Errors rect,fig:Errors sin2} we compare  $\hat U_{n}$ to $\hat U_\text{num}$ with the drive amplitude set to $\Omega=\Omega_2$, but it is worth noting that the results are similar for $\Omega=\Omega_4$.

In the following, we provide a comparison between the infidelity calculated using different propagators as a function of various parameters, as shown in \cref{fig:Errors rect}.
As a general observation, in \cref{fig:Errors rect}, it can be seen that there is a strong agreement between the curve for $\hat U_4$ and the curve for $\hat U_\text{num}$.
The dependence of the infidelity on the initial average motional mode population is plotted in \cref{fig:Errors rect nbar}.
At low temperatures, the curves calculated with $\hat U_2$ and $\hat U_3$ predict lower infidelity than those calculated with $\hat U_4$ and $\hat U_\text{num}$, highlighting their lack of predictive power. 
However, at higher temperature, all predicted infidelities increase monotonically and nearly converge. 
The infidelity is almost linear in the motional mode population $\bar n$, as expected from the terms which are linear in $n$ in \cref{tab:Error budget}.

\Cref{fig:Errors rect K} illustrates the predicted infidelity as a function of the dimensionless gate time $K=\frac{\nu T}{2\pi}$, indicating a threshold around $K=100$ below which the predicted infidelity increases significantly as $K$ decreases.
For gate times longer than $K\approx 100$, the curves representing $\hat U_3$ and $\hat U_4$ converge, whereas for gate times shorter than $K\approx100$, the predicted infidelities diverge, as expected when the coherent errors increase and the ratio of $\Omega$ to the other frequencies in the system is increased. In particular, the predicted infidelities calculated with $\hat U_4$ and $\hat U_\text{num}$ increase significantly as $K$ decreases. Meanwhile, the infidelities calculated with the lower orders of the Magnus expansion $\hat U_2$ and $\hat U_3$ remain nearly horizontal, indicating an inability to capture the full dynamics at short times, but the curve calculated with $\hat U_2$ lies an order of magnitude below the infidelity calculated with $U_3$. 

\Cref{fig:Errors rect eta} depicts the predicted infidelity as a function of the Lamb-Dicke parameter $\eta$. The curve calculated with $\hat U_2$ lies approximately an order of magnitude lower than the curve calculated with $\hat U_3$.
The curves for $\hat U_4$ and $\hat U_\text{num}$ do approach the curve for $\hat U_3$ around $\eta = 0.3$ but this is in the higher error regime; on the other hand, they diverge significantly at smaller values of $\eta$, once again disagreeing strongly with the standard predictions. 
Notably, the curves calculated with $\hat U_2$ and $\hat U_3$ increase monotonically with $\eta$, while the curves calculated with $\hat U_4$ and $\hat U_\text{num}$ have a minimum around $\eta=0.2$.
This indicates that making the Lamb-Dicke parameter as small as possible does not improve the infidelity of the \moso{}. Instead, there is an optimal value for the Lamb-Dicke parameter for the \moso{}.

\subsection{Discussion}
In \cref{fig:Energy level diagram}, the terms of Section~\ref{sec:Error terms} are visually represented. 
Through our analysis in the previous sections, it has become clear that the second order of the Magnus expansion, $\hat Z_2$, is not the only term that significantly impacts the gate fidelity.
In fact, terms up to and including the fourth order $\hat Z_4$ contribute significantly to the fidelity.
Although higher orders in the Magnus expansion may in principle be relevant as well, \cref{sec:Numerical Results rect} demonstrates that they are typically not necessary for commonly used parameters.

In a model that accounts only for the second order of the Magnus expansion, one might be tempted to choose a small value for the Lamb-Dicke parameter $\eta$ to minimize errors related to the Lamb-Dicke expansion.
However, as demonstrated in \cref{tab:Error budget}, reducing $\eta$ decreases the error in $\hat Z_2$ and $\hat Z_3$, but also increases the error term in $\hat Z_4$. This indicates the existence of a sweet spot for $\eta$, as shown in \cref{fig:Errors rect eta}.
The fourth order term $\hat Z_4$ leads, due to the additional term proportional to $\hat J_y^2$, to a shift in the optimal drive amplitude without significantly impacting the fidelity compared to $\hat Z_3$. Meanwhile, the contribution of $\hat Z_3$, compared to $\hat Z_2$, increases the infidelity, as illustrated in  \cref{tab:Error budget,fig:Errors rect Omega}. 

The temperature dependence of the error terms, as presented in \cref{tab:Error budget}, reveals that certain terms are proportional to $\bar n$, consistent with the trend observed in \cref{fig:Errors rect nbar}.

According to \cite{sorensen_entanglement_2000}, one can neglect $\hat Z_3$, because the error in $\hat Z_2$ is $\bigO{n\eta^2}$, while the error in $\hat Z_3$ is $\bigO{\eta^2}$. They argue that deviations of the Lamb-Dicke approximation are typically caused by large values of $n$ rather than $\eta$, making $\hat Z_3$ insignificant compared to $\hat Z_2$. 
Nevertheless, this perspective fails to consider that error terms related to higher sideband orders could have lower orders in the Lamb-Dicke expansion, as seen in the case of the second sideband order of $\hat Z_3$ (\cref{eq:Z3 m2}).
Moreover, due to improved cooling methods \cite{mehta_integrated_2020} since then, it is increasingly crucial to have expressions that remain valid at low temperatures.

\section{Pulse Shaping}\label{sec:Pulse Shaping}

\begin{figure*}[ht]
    \centering
    \begin{subfigure}[h]{0.32\textwidth}
    \centering
    \includegraphics[width=1\linewidth]{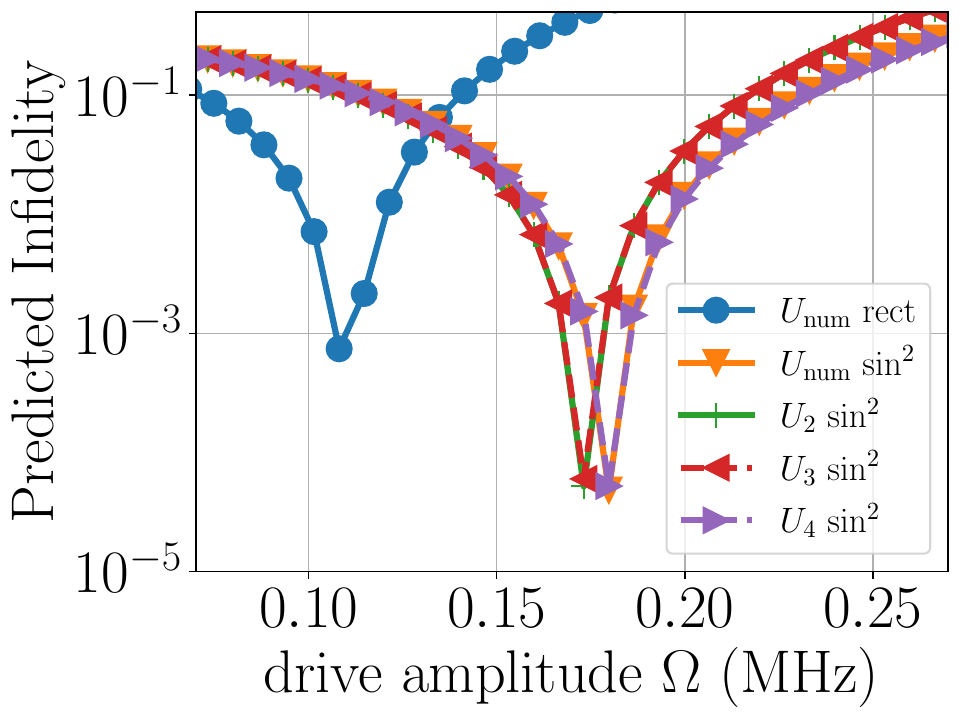}
      \caption{}
    \label{fig:Errors sin2 Omega}
    \end{subfigure}
    \begin{subfigure}[h]{0.32\textwidth}
        \centering
        \includegraphics[width=1\linewidth]{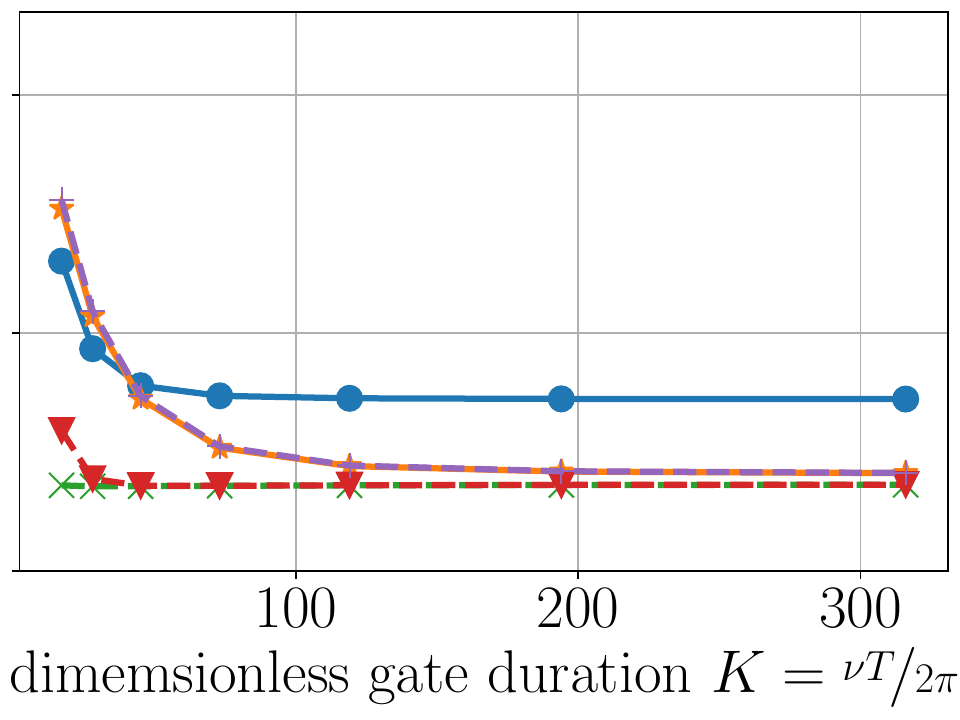}
      \caption{}
        \label{fig:Errors sin2 K}
    \end{subfigure}
    \begin{subfigure}[h]{0.32\textwidth}
        \centering
        \includegraphics[width=1\linewidth]{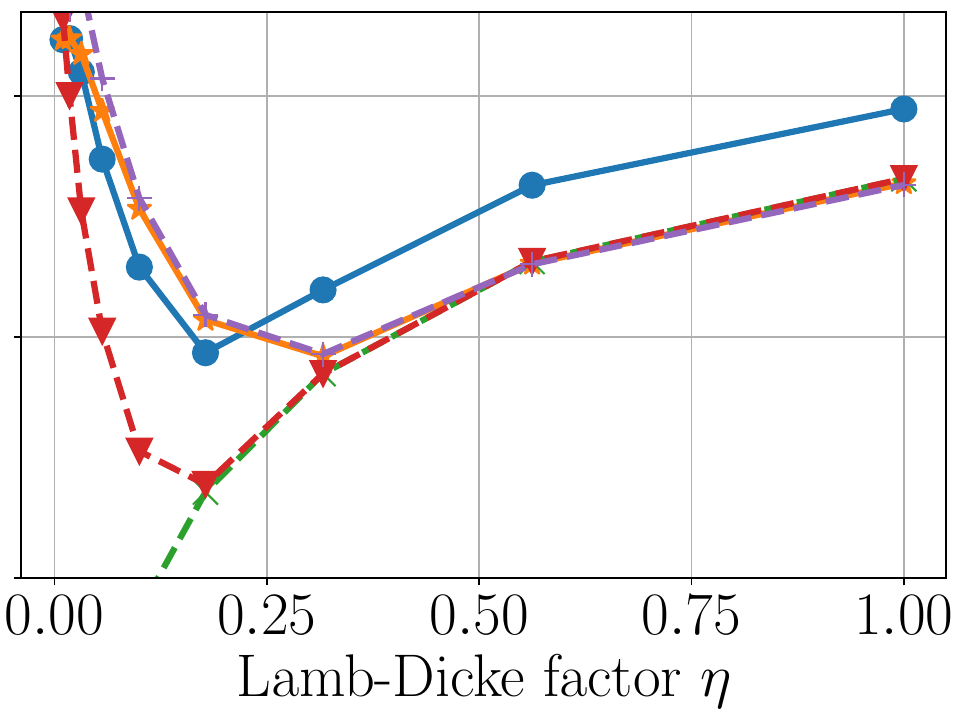}
      \caption{}
        \label{fig:Errors sin2 eta}
    \end{subfigure}
	\caption{Predicted average infidelity of the \moso{} $\sin^2$-pulses and for comparison the numerical predicted infidelity for a rectangular pulse as a function of drive amplitude $\Omega$, dimensionless gate duration $K$ and Lamb-Dicke factor $\eta$, respectively. 
	The parameters are (if they are not varied): $\eta=0.18$, $K-L=\frac{\nu-\delta}{2\pi}T=3$, $K=28$, $\nu/(2\pi)=\SI{0.1}{\mega\hertz}$, $\bar n=\num{2e-2}$. 
The data for \cref{fig:Errors sin2 K,fig:Errors sin2 eta} is calculated using $\Omega = \Omega_{2,\text{rect}}$, \cref{eq:Omega opt U2}, for the rectangular pulse shape and $\Omega = \SI{0.173}{\mega\hertz}$ for the $\sin^2$-pulse shape.}
    \label{fig:Errors sin2}
\end{figure*}

So far, we have scrutinized the problem for a rectangular drive amplitude. However, the analysis in Section~\ref{sec:Resonance Conditions} is valid for arbitrary finite-window, smooth pulse shapes, with small modifications as given in \cref{sec:Resonance integral shaped}.
Amplitude modulation experiments on the \moso{} have already been performed \cite{zarantonello_robust_2019,steane_pulsed_2014}, but to the authors' knowledge, no theoretical investigation has been realized so far.
A straightforward candidate for a smooth pulse is a Gaussian pulse, but since it is not finite in time, we fall back on a similar, and well-behaved pulse shape of the form
\begin{equation}
    \Omega(t) = \Omega \sin[2](\frac{\pi}{T}t).
\end{equation}
The $\sin^2$ pulse shape has also a narrower bandwidth (i.e.~smaller maximal value of $M$), when compared to a truncated Gaussian pulse. This is potentially preferable since fewer unwanted transitions may be driven, and in general more compatible with keeping track of all the resonance conditions, discussed in Section~\ref{sec:Resonance Conditions}.
Since the analytical results for a $\sin^2$-pulse are lengthy, we only sketch some terms in this section and give further details in \cref{sec:Resonance integral shaped} and \cref{sec:Magnus terms shaped}.

\subsection {Analytic expressions}\label{sec:Analytical expressions sin2}

The 2nd order of the Magnus expansion for a $\sin^2$-pulse is of the form
\begin{equation}\label{eq:Z2 sin2}
\begin{aligned}
            \hat Z_{2,\sin^2} = \frac{K\Omega^2T^2\eta^2}{\pi}&\left(\frac{p_y}{q_y}\left(1-(2n+1)\eta^2\right)\hat J_y^2 \right.\\&\left.+ \frac{p_x}{q_x}\left(2n+1\right)\eta^2 \hat J_x^2 \right),
\end{aligned}
\end{equation}
where $p_{x,y}$ and $q_{x,y}$ are polynomials in $K$ and $L$, defined in \cref{sec:Magnus terms shaped}. From this expression, one can derive an optimal drive amplitude
    \begin{equation}\label{eq:Omega opt LD sin2}
        \Omega_{\text{LD},\sin^2} = \frac{\pi}{\eta T \sqrt{2K}}\sqrt{\frac{q_y}{p_y}}.
    \end{equation}
    The 3rd order of the Magnus expansion is of the form
    \begin{equation}
        \label{eq:Z3 sin2}
        \hat Z_{3, \sin^2} = \frac{K^2\Omega^3T^3\eta^5}{\pi^2}\frac{p_3}{q_3}\hat J_y \left(\hat a+\hat a^\dagger\right),
    \end{equation}
    where $p_3$ and $q_3$ are again polynomials in $K$ and $L$.
    
    One can show that the resulting error terms at the optimal drive amplitude are smaller than the error terms for a rectangular pulse, which have the same form, but different prefactors. This is, because the prefactor $\frac{p_3}{q_3}$ is smaller in the case of a $\sin^2$-pulse than the respective one for a rectangular pulse given in \cref{tab:Error budget}, for most realistic choices of $K$ and $L$.

\subsection{Numerical Comparison}\label{sec:Numerical Results sin2}

We now compare the average infidelity, $1-\mathcal{F}$, using a rectangular pulse $\Omega(t)= \Omega$ to the infidelity using a $\sin^2$-pulse $\Omega(t) = \Omega\sin[2](\frac{\pi}{T}t)$.
Explicitly, we compare $\hat U_n$ for $n=2,\cdots,4$, to $\hat U_\text{num}$ using a $\sin^2$-pulse, while also comparing to $\hat U_\text{num}$ using a rectangular pulse. The propagators $\hat U_n$, $\hat U_\text{num}$ are calculated with \cref{eq:def Un,eq:def Unum}, respectively.

Comparing the predicted infidelities calculated with $\hat U_\text{num}$ as a function of $\Omega$ for a rectangular and a $\sin^2$-pulse, one can see in \cref{fig:Errors sin2 Omega} that the optimal (highest reached) drive amplitude differs for different pulse shapes, but that the minimal predicted infidelity for a $\sin^2$-pulse is smaller than for a rectangular pulse.
Because the optimal drive amplitudes differ, in \cref{fig:Errors sin2 K,fig:Errors sin2 eta} the fidelities are plotted at their respective minimal drive amplitude $\Omega_2$.

In \cref{fig:Errors sin2 K}, the infidelity is shown as a function of the dimensionless gate duration $K$. In the low infidelity regime, the shaped pulses outperform the square pulses considerably, with both the full numerics and the fourth Magnus order closely agreeing.
The graph also shows that inclusion of orders higher than three in the Magnus expansion (described by $\hat U_4$) results in an increase of the predicted infidelity as the value of K decreases, but only below a certain threshold that is dependent on $\eta$.
At dimensionless gate times longer than $K\approx 50$, the calculation using a $\sin^2$-pulse predicts a lower infidelity, but at gate times shorter than $K\approx 50$, the rectangular pulse yields a lower infidelity. One can see that for the $\sin^2$-pulse, except for short gate times, there is  a close agreement between the curve calculated with $\hat U_2$ and with $\hat U_3$, which implies that pulse shaping can probably compensate for the off-diagonal errors in $\hat Z_3$, unlike in \cref{fig:Errors rect K}.

For the Lamb-Dicke parameter $\eta$, we see once again that the contributions of the higher orders in the  Magnus expansion lead correctly to the existence of an optimal value for $\eta$, while instead the terms $\hat Z_2$ and $\hat Z_3$ decrease with decreasing $\eta$.
The data in \cref{fig:Errors sin2 eta} is not detailed enough to tell which pulse shape in general leads to a smaller minimal predicted infidelity. For Lamb-Dicke parameters larger than the minimal value, the $\sin^2$-pulse yields a much lower infidelity, but for small $\eta$, the rectangular pulse is slightly better.
The behavior of the infidelity as a function of $\eta$ is similar for rectangular pulses and $\sin^2$-pulses.

\subsection{Discussion}\label{sec:Discussion}
\Cref{fig:Errors sin2} illustrates the potential for using shaped pulses to further reduce the error in the \moso{}. At a quantitative level, we see that while in the previous plots parameter searches limited infidelities to the $10^{-3}$ range, there are promising parameter regimes with $\sin^2$-pulses that lead to errors below $10^{-4}$. This can be explained mathematically, through the Magnus expansion, as arising out of generally smaller prefactors in front of the same error operators terms.

At a qualitative level, the behavior of the infidelity for varying parameters for a $\sin^2$-pulse is similar to the behavior of the infidelity for a rectangular pulse. Moreover, for certain parameter regimes, the rectangular pulses actually outperform the $\sin^2$-strategy.  This is consistent both with conventional wisdom on pulse shaping and the nuances of the \moso{}. Indeed, square pulses are known to allow spectral hole burning through their characteristic $\text{sinc}$ spectrum, while smooth pulses offer the possibility of reducing bandwidth overall and thereby avoid entire ranges of off-resonant errors. In the case of the \moso{}, the spectral conditions on beat-note indices not only reduce errors at low Magnus orders, but in certain cases also eliminate certain other terms, and so for a fortuitous choice of parameters it may indeed offer the best strategy. Smooth shaping allows nonetheless generally a safer option, and indeed seem to perform optimally in the best case. Lastly, the latter can be combined more easily with other control theoretic methods to potentially reduce errors even further.

\section{Summary}\label{sec:Summary}
We showed that calculating the Magnus expansion only up to the second order $\hat Z_2$ is not sufficient to adequately describe the \moso{}. 

While the Lamb-Dicke expansion is indeed a valid approximation for the first nonvanishing Magnus order $\hat Z_2$, we show that higher orders of the Magnus expansion, and explicitly $\hat Z_3$ and $\hat Z_4$, cannot be neglected in general.
The contribution of the third order $\hat Z_3$ increases significantly the infidelity, being of the same order $\bigO{\eta^4}$ in $\eta$ as the Lamb-Dicke error terms in $\hat Z_2$.
Moreover, the fourth order $\hat Z_4$ is even of the same order $\bigO{\eta^2}$ in $\eta$ as the gate term in $\hat Z_2$, leading to a significant shift in the optimal drive amplitude. Due to being of different order in $\Omega$, a sweet spot for the Lamb-Dicke coefficient $\eta$ also emerges.  

We give predictive expressions for the optimal drive amplitude (up to experimental calibration), and show qualitatively that an optimal $\eta$ may be found, e.g. through numerical calibration.
Due to the contribution of the third order in the Magnus expansion $\hat Z_3$, the \moso{} fidelity is inherently limited, but with analytical expressions for leading terms of the first four Magnus terms, this opens potential avenues to suppress certain terms.  For example, we show that an amplitude with envelope $\sin[2](\frac{\pi}{T}t)$ decreases the contribution of the third order of the Magnus expansion. The presented framework also opens new paths for application of optimal control methods on the \moso{}, such as the use of DRAG pulses \cite{theis_counteracting_2018}. It also makes it straightforward to investigate the effect of multichromatic beams.

\begin{acknowledgments}
This work was supported by the Helmholtz Validation Fund project HVF-00096 and under Horizon Europe programmes HORIZON-CL4-2021-DIGITALEMERGING-02-10 via the project 101080085 (QCFD) and HORIZON-CL4-2022-QUANTUM-02-SGA via the project 101113690 (PASQuanS2.1). 
\end{acknowledgments}

\bibliography{MyLibraryBibtex}
\appendix

\section{General expressions for the resonance integral with shaped pulses}\label{sec:Resonance integral shaped}

\subsection{First order} \label{sec:first order shaped}
The general expression for the first order of the Magnus expansion \cref{eq:Z1} is
\begin{equation}
    \label{eq:Z1 shaped}
	\hat Z_1 = \Omega\sum_{M,m=-\infty}^\infty\sum_{\mu=\pm1} c_M \underbrace{\frac{e^{i\left(\frac{2\pi}{T}M+m\nu+\mu \delta\right)T}-1}{i\left(\frac{2\pi}{T}M+m\nu+\mu \delta\right)}}_{\text{resonance condition}}\hat J_m\hat A_m.
\end{equation}
Using the beat-note index
\begin{equation}
\label{eq:N shaped}
    N(M,m,\mu) := M + \underbrace{\frac{\nu T}{2\pi}}_K m + \underbrace{\frac{\delta T}{2\pi}}_L\mu
\end{equation} 
the Hamiltonian can be written more compactly as
\begin{equation}
    \label{eq:Hamilton with integer shaped}
	\hat H(t) = \hbar\Omega \sum_{M,m=-\infty}^\infty\sum_{\mu=\pm1} c_M  e^{i\frac{2\pi}{T} ( \underbrace{M + K m + L\mu}_{N})t}\hat J_m\hat A_m.
\end{equation}

\subsection{Second Order} \label{sec:second order shaped}
The second order of the Magnus expansion \cref{eq:Z2 general} with the resonance integral \cref{eq:Int 2} and the beat-note indices
\begin{equation}\label{eq:Nj shaped}
    N_j = M_j + m_j K + \mu_j L
\end{equation}
can be simplified further if the two-photon resonance condition
\begin{equation}
\label{eq:multipartite resonance condition shaped}
\begin{split}
    M_1+m_1K+\mu_1 L+M_2+m_2K+\mu_2 L=0\\ 
    \iff (M_1=-M_2) \land (m_1=-m_2) \land (\mu_1=-\mu_2)
\end{split}
\end{equation}
is satisfied (see \cref{sec:Magnus term condition}).

The general form of the form factors \cref{eq:Z2 coeffs} is
\begin{equation}
    \label{eq:Z2 coeffs shaped}\nonumber
    \begin{split}
    d_{x,y}^{(n)} = \mp\frac{\Omega^2T^2}{2\pi} e^{-\eta^2} &\sum_{m \text{ even, odd}}\sum_{M\mu} \frac{|c_{M}|^2 }{M+mK+\mu L} (-\eta^2)^{|m|} \\
    & \left(L_{\min(n,n-m)}^{|m|}(\eta^2) \right)^2\frac{\min(n,n-m)!}{\max(n,n-m)!}.
    \end{split}
\end{equation}

\section{Choice of fidelity}\label{sec:choice of fidelity}
The Bell state fidelity, for an initial state $\rho_0$ and a target qubit state $\ket{\psi_\text{target}}$,
\begin{equation}\label{eq:Bell fidelity}
		\mathcal{F}_\text{Bell} = \mel{\psi_\text{target}}{\text{Tr}_\text{motion}\{\hat U(T) \rho_0 \hat U^\dagger(T)\}}{\psi_\text{target}}
\end{equation} 
is often used in experiments \cite{sorensen_entanglement_2000,mehta_integrated_2020,mehta_fast_2019,clark_high-fidelity_2021,ballance_high-fidelity_2016,tinkey_quantum_2021,gaebler_high-fidelity_2016} and therefore allows for straightforward comparison. The main advantage of the metric is that it is a good measure of (maximally) achievable entanglement, while at the same time ignoring certain relative phases that can otherwise be corrected, e.g.,~with single qubit gates. 
Also, the Bell state fidelity allows for straightforward calculation of the fidelity for Hamiltonians of the form
\begin{equation}
     \sum_n\sum_{j=x,y,z}\left(d_j^{(n)}\hat J_j^2\right)\otimes \dyad{n}{n}.
	 \label{eq:eff Hamiltonian diag}
\end{equation}
Since $\hat J_x^2$ and $\hat J_y^2$ are qubit entangling operators, this is an entangling operation which does not affect the state of the motional mode. It yields for an initial qubit state $\ket{\psi_0} = \ket{00}$, an entangled qubit target state $\ket{\psi_t}=\frac{1}{\sqrt{2}}\left(\ket{00}+e^{i\phi}\ket{11}\right)$, and an initial motional mode state $\sum_{n=0}^\infty P_n \dyad{n}{n}$, where $P_n$ is the initial probability for the motional mode to be in state $\ket{n}$, the Bell state fidelity
\begin{equation}\label{eq:Bell fidelity sin dx dy}
    \mathcal{F}_\text{Bell} = \frac{1}{2} \left(1-\sum_n P_n\sin(\phi)\sin(d_x^{(n)}-d_y^{(n)})\right).
\end{equation}
 Hence the optimal fidelity $\mathcal F_\text{Bell}=1$ for a target state $\ket{\psi_t} = \frac{1}{\sqrt{2}}\left(\ket{00} - i\ket{11}\right)$, with rotation angle $\phi=-\frac{\pi}{2}$, is obtained if
 \begin{equation}\label{eq:opt Fid cond}
     \sum_n P_n\sin(d_x^{(n)}-d_y^{(n)}) =  1
 \end{equation}
is fulfilled. 
This equation can be solved numerically, but to get an analytical solution, further approximations must be made.
Assuming $P_n$ decays quickly with $n$ then one can, e.g., assume a thermal distribution $P_n = \frac{(\bar n)^n}{(\bar n +1)^{n+1}}$, with small average phonon number $\bar n\ll1$.
Then, the terms for $n>0$ can be neglected, and the condition for optimal fidelity becomes $\sin(d_x^{(0)}-d_y^{(0)}) =  1 $, and thus
 \begin{equation}\label{eq:opt Fid cond n0}
     d_x^{(0)}-d_y^{(0)}=  \frac{\pi}{2}.
 \end{equation}

This equation is valid for the leading terms of the effective Hamiltonian for the \moso{}. However, the Bell state fidelity requires fixing specific initial and target states.

Another measure which allows quantification of the fidelity in quantum computing is the average (gate-overlap) fidelity 
\begin{equation}\label{eq:weighted average fidelity}
		\mathcal{F}_\text{av} = \frac{1}{4}\left|\text{Tr}_\text{qubits}\left\{\sum_{n=0}^\infty P_n\mel{n}{\hat U(T) \hat U_\text{target}^\dagger}{n}\right\}\right|,
\end{equation} 
with $\hat U_\text{target} = \text{exp}(i \varphi \hat J_y^2)$, $\varphi=\frac{\pi}{2}$.
It yields nearly the same results as the Bell fidelity if the trace over the motional modes is weighted by the Boltzmann distribution, $P_n = \frac{\bar n^n}{(\bar n+1)^{n+1}}$ ensuring that performance is measured accordingly for more likely initial states.

The average fidelity is a more standard measure for quantum information settings \cite{bermudez_assessing_2017}, because it measures how close a gate is to the target gate, which is useful for implementation of quantum algorithms.

In order to calculate $\Omega_\text{LD}$ and $\Omega_4$, the average fidelity can be used directly.

Comparison of the target propagator and the effective Hamiltonian \cref{eq:eff Hamiltonian diag} shows that the average fidelity needs small $d_x$ and $d_z$ and additionally 
\begin{equation}
		d_y = -\frac{\pi}{2}.
		\label{eq:average fidelity dy}
\end{equation}
To calculate the impact of the $\hat J_x^2$-term, the Bell fidelity is quite handy, since \cref{eq:Bell fidelity sin dx dy} is concise expression containing both $d_x$ and $d_y$. The resulting $\Omega_2$ yields a small correction to $\Omega_\text{LD}$.

\begin{figure}[ht]
    \centering
	\includegraphics[width=1\linewidth]{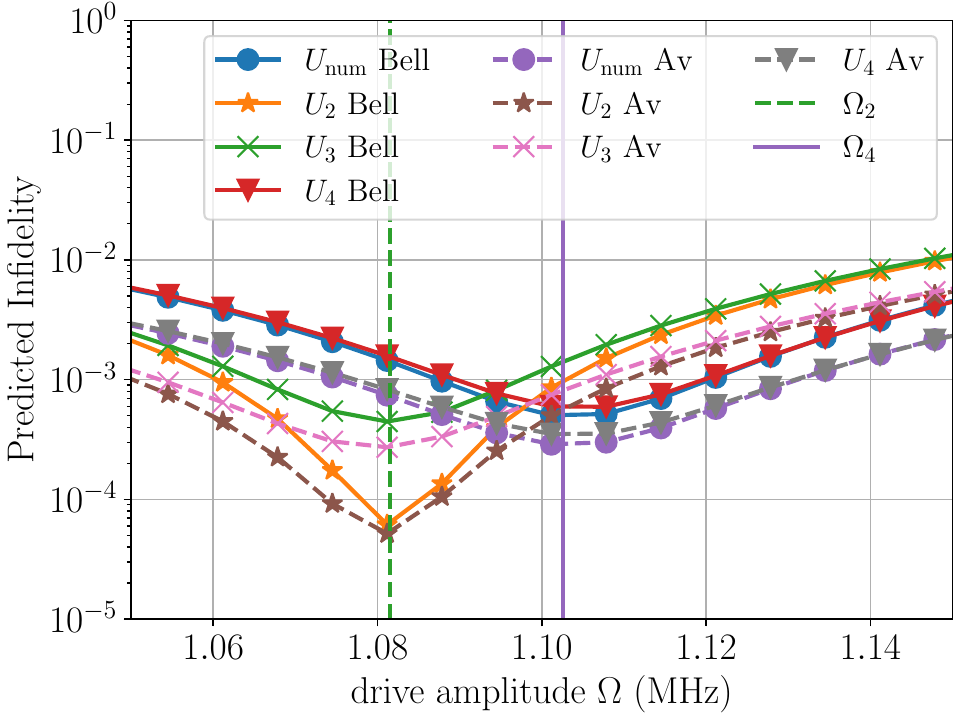}
	\caption{Comparison of the results for average and Bell state infidelity, using the same parameters as in \cref{fig:Errors rect}}
    \label{fig:Errors rect fidelities}
\end{figure}

Although the Bell state fidelity requires specification of the involved states and the average fidelity requires a specific target gate, the results are nonetheless quite similar: For the parameters used in \cref{fig:Errors rect} at $\Omega=\Omega_2$, the Bell state infidelity is $1-\mathcal{F}_\text{Bell}(\hat U_\text{num}) = \num{1.3e-3}$ and the average infidelity $1-\mathcal{F}_\text{av}(\hat U_\text{num}) = \num{0.67e-3}$. 
At $\Omega=\Omega_4$ the infidelities are $1-\mathcal{F}_\text{Bell}(\hat U_\text{num}) = \num{4.3e-4}$ and $1-\mathcal{F}_\text{av}(\hat U_\text{num}) = \num{2.4e-4}$ for the Bell state infidelity and the average infidelity, respectively. A comparison over the range of amplitudes is shown in \cref{fig:Errors rect fidelities}.

Beyond this discussion, the error budget terms presented in \cref{tab:Error budget} provide the precise Hamiltonian description and do not hinge on a particular fidelity metric. As such, they are relevant to any particular cost function that is used.

\section{Conditions on the variables for the analytical terms}\label{sec:Magnus term condition}
The terms in \cref{tab:Error budget} assume that the detuning avoids certain sideband ranges, which is justified as follows.
In order to simplify the expression for $\hat Z_2$, \cref{eq:Z2 general}, we have assumed in \cref{eq:multipartite resonance condition} that 
\begin{equation}
		N_1+N_2 = 0 \iff (m_1+m_2=0) \land (\mu_1+\mu_2 = 0)
\end{equation}
To drive a single motional mode the laser detuning must be close to the motional mode frequency $K-L\ll K$. 
If $\mu_1+\mu_2\neq 0$, then the resonance condition $N_1+N_2=0$ can be rewritten as
\begin{equation}
		1+\frac{m_1+m_2}{\mu_1+\mu_2} = \frac{K-L}{K} \ll 1.
\end{equation}
Checking the condition for the allowed values of $\mu_1+\mu_2$ yields:
\begin{enumerate}
		\item $1 - \frac{m_1+m_2}{2} \ll1$ for $\nu_1+\nu_2=-2$ This can only be fulfilled for $m_1+m_2=1$ and yields the resonance condition $K=2L$.
		\item $1-(m_1+m_2\ll1$ does not yield a result for integer $m_1,m_2$.
		\item Nor does $1+(m_1+m_2)\ll1$.
		\item $1 + \frac{m_1+m_2}{2} \ll 1$ can be fulfilled for $m_1+m_2=-1$ and yields the resonance condition $K=2L$.
\end{enumerate}
Thus, as long as $K-L \ll K$ and specifically $K\neq2L$, condition \cref{eq:multipartite resonance condition} is valid.

One can make similar assumptions for the other orders of the Magnus expansion.
So, for simplicity, to calculate the terms in \cref{tab:Error budget}, we excluded at each $k$-th order of the Magnus expansion terms where $jK=lL$ if one of the following conditions on $\frac{j-l}{l}=\frac{K-L}{K}$ is true:
\begin{enumerate}
        \item $\frac{l-j}{l} \leq 0$, which is false, because by construction $K>0$ and $K-L>0$. 
        \item $\frac{l-j}{l} \geq \frac{1}{km_\text{max}}$, where $m_\text{max}$ is the maximum value of $m$ which is included in the calculation of the Magnus expansion (for our calculations $m_\text{max}=3$).
\end{enumerate}
In most cases where $\frac {K-L}{K}\ll 1$, the second condition is wrong as well.
For small $K$ and larger $K-L$ (e.g.\ $K=28$ and $K-L=3$), one has to make sure that there is no $j\leq k m_\text{max}$ and $|l|\leq k$ such that $jK=lL$ (which is fine for $K=28$ and $K-L=3$).

\section{Full expressions for Magnus expansion of the shaped pulse}\label{sec:Magnus terms shaped}
The full expressions for the Magnus expansion for a $\sin^2$-pulse are the following: For the 2nd order of the Magnus expansion, \cref{eq:Z2 sin2},
the coefficients are 
\begin{align}
    p_y&=3(K^{2}-L^{2})^2-4(5K^{2}+3L^{2}-8),\\
    q_y&=8\left(K^{2}-L^{2}\right)\left(\left(K-L\right)^{2}-4\right)\left(\left(K+L\right)^{2}-4\right)
\end{align}
for the first sideband, and
\begin{align}
    p_x&=8\left(6K^{4}-3K^{2}L^{2}-10K^{2}+4\right)+3\left(L^{4}-4L^{2}\right), \\
    q_x&=8\left(4K^{2}-L\right)\left(\left(2K-L\right)^{2}-4\right)\left(\left(2K+L\right)^{2}-4\right)
\end{align}
for the second sideband.
    
The optimal drive amplitude (taking only the leading Lamb-Dicke order and the first sideband into account) is shown in \cref{eq:Omega opt LD sin2}. 

The coefficients for the 3rd order of the Magnus expansion, \cref{eq:Z3 sin2}, are
\begin{align}   
    p_3&=\left(K^{2}+3L^{2}-4\right)\left(3\left(K^{2}-L^{2}\right)^2-4\left(5K^{2}+3L^{2}-8\right)\right), \\
    q_3&=8\left(K^{2}-L^{2}\right)^{2}\left(\left(K-L\right)^{2}-4\right)^{2}\left(\left(K+L\right)^{2}-4\right)^{2}.
\end{align}

\end{document}